\documentclass{LMCS}

\def\doi{8 (3:12) 2012}
\lmcsheading%
{\doi}
{1--27}
{}
{}
{Nov.~\phantom06, 2009}
{Aug.~13, 2012}
{}

\usepackage{hyperref,enumerate}

\usepackage{proof}
\usepackage{amsmath}
\usepackage{subfigure}
\usepackage{epsf}
\usepackage{graphics}
\usepackage{wrapfig}
\usepackage{mathrsfs}
\usepackage{url}
\usepackage{calc}
\usepackage{amsmath,amssymb,amsfonts,mathrsfs,latexsym,stmaryrd}
\usepackage[all]{xy}

\newcommand{\funone}{\mathbf{f}}
\newcommand{\funtwo}{\mathbf{g}}
\newcommand{\funthree}{\mathbf{h}}
\newcommand{\funfour}{\mathbf{p}}
\newcommand{\conone}{\mathbf{c}}

\newcommand{\patone}{s}
\newcommand{\pattwo}{r}
\newcommand{\patthree}{q}
\newcommand{\seqpone}{\alpha}
\newcommand{\seqptwo}{\beta}
\newcommand{\varone}{x}
\newcommand{\vartwo}{y}
\newcommand{\varthree}{z}
\newcommand{\varfour}{w}
\newcommand{\varfive}{f}
\newcommand{\varsix}{g}
\newcommand{\varseven}{h}
\newcommand{\lambdaone}{M}
\newcommand{\lambdatwo}{N}
\newcommand{\lambdathree}{L}
\newcommand{\lambdafour}{P}
\newcommand{\lambdafive}{Q}
\newcommand{\valueone}{V}
\newcommand{\valuetwo}{W}
\newcommand{\valuethree}{Z}
\newcommand{\termone}{t}
\newcommand{\termtwo}{u}
\newcommand{\termthree}{v}
\newcommand{\termfour}{w}
\newcommand{\termfive}{d}
\newcommand{\termsix}{e}
\newcommand{\termseven}{f}
\newcommand{\cltermone}{X}
\newcommand{\cltermtwo}{Y}
\newcommand{\sigone}{\Sigma}

\newcommand{\vsone}{V}

\newcommand{\ordone}{\alpha}

\newcommand{\labelone}{\delta}

\newcommand{\rootone}{r}
\newcommand{\roottwo}{s}
\newcommand{\verone}{v}

\newcommand{\tgone}{G}
\newcommand{\tgtwo}{H}
\newcommand{\tgthree}{J}
\newcommand{\tgfour}{K}
\newcommand{\tgfive}{I}
\newcommand{\rrone}{\rho}

\newcommand{\homone}{\varphi}

\newcommand{\nat}[1]{\lceil #1\rceil}
\newcommand{\rewrite}[1]{\stackrel{#1}{\longrightarrow}}

\newcommand{\appTRS}{\mathbf{app}}
\newcommand{\cappTRSW}{\mathbf{capp}}
\newcommand{\constr}[2]{\mathbf{c}_{#1,#2}}
\newcommand{\LambdatoTRS}[1]{[#1]_{\Phi}}
\newcommand{\TRStolambda}[1]{\langle{#1}\rangle_{\Lambdaterms}}
\newcommand{\TRStolambdaI}[1]{\langle\!\langle #1\rangle\!\rangle_{\Lambdaterms}}
\newcommand{\TRStolambdaII}[1]{[#1]_{\Lambdaterms}}
\newcommand{\TRSonetolambdaI}[1]{\langle\!\langle #1\rangle\!\rangle_{\Lambdaterms}}
\newcommand{\TRSonetolambdaII}[1]{[#1]_{\Lambdaterms}}
\newcommand{\TRSonetolambdaIII}[1]{\langle\!|#1|\!\rangle_{\Lambdaterms}}
\newcommand{\TRSWtolambda}[1]{\langle #1\rangle_{\Lambdaterms}}
\newcommand{\TRStoGRS}[1]{[#1]}
\newcommand{\TRStoGRSW}[1]{[#1]_{\Xi}}
\newcommand{\GRStoTRS}[1]{\langle #1\rangle}
\newcommand{\LambdatoTRSW}[1]{[#1]_{\Psi}}
\newcommand{\LambdatoTRSWaux}[1]{\{#1\}_{\Psi}}
\newcommand{\errorterm}{\bot}
\newcommand{\arity}[1]{\mathit{ar}(#1)}
\newcommand{\dsize}[1]{||#1||}

\newcommand{\Variables}{\Upsilon}
\newcommand{\Lambdaterms}{\Lambda}

\newcommand{\Functions}[1]{\Sigma_{#1}}
\newcommand{\Rules}[1]{\mathcal{R}_{#1}}
\newcommand{\Graphs}[1]{\mathcal{G}_{#1}}
\newcommand{\TRS}{\Phi}
\newcommand{\TRSterms}{\mathcal{T}(\Phi)}
\newcommand{\TRSvarterms}{\mathcal{V}(\Phi,\Variables)}
\newcommand{\TRSconterms}{\mathcal{C}(\Phi)}
\newcommand{\TRSone}{\Xi}

\newcommand{\TRStermsp}[1]{\mathcal{T}(#1)}
\newcommand{\TRSpatsp}[1]{\mathcal{P}(#1,\Variables)}
\newcommand{\TRSvartermsp}[1]{\mathcal{V}(#1,\Variables)}
\newcommand{\TRScontermsp}[1]{\mathcal{C}(#1)}
\newcommand{\GRS}{\Theta}
\newcommand{\GRSW}{\Xi}
\newcommand{\CtoCG}[1]{[#1]}
\newcommand{\CGtoC}[1]{\langle #1\rangle}
\newcommand{\TRSW}{\Psi}
\newcommand{\TRSWterms}{\mathcal{T}(\Psi)}
\newcommand{\TRSWvarterms}{\mathcal{V}(\Psi,\Variables)}
\newcommand{\TRSWconterms}{\mathcal{C}(\Psi)}
\newcommand{\domain}[1]{\mathit{dom}(#1)}
\newcommand{\srrone}{\Rules{}}

\newcommand{\sgrone}{\Graphs{}}
\newcommand{\Time}[1]{\mathit{Time}_v(#1)}
\newcommand{\Timew}[1]{\mathit{Time}_h(#1)}

\newcommand{\TMone}{\mathcal{M}}

\newcommand{\rewrTRS}{\rightarrow}
\newcommand{\rewrTRSW}{\rightarrow}
\newcommand{\rewr}[1]{\rightarrow_{#1}}
\newcommand{\rewrlambdav}{\rightarrow_v}
\newcommand{\rewrlambdah}{\rightarrow_h}
\newcommand{\rewrgraph}{\rightarrow}

\newcommand{\subgr}[2]{#1\downarrow #2}

\newcommand{\cod}[1]{\ulcorner #1\urcorner}

\newcommand{\N}{\mathbb{N}}

\newcommand{\FV}[1]{\mathtt{FV}(#1)}
\newcommand{\length}[1]{|#1|}
\newcommand{\plength}[2]{|#1|_{#2}}

\newtheorem{proposition}{Proposition}

\newenvironment{varitemize}
{
\begin{list}{\labelitemi}
{\setlength{\itemsep}{0.0mm}
 \setlength{\topsep}{0.0mm}
 \setlength{\parindent}{0.0mm}
 \setlength{\parskip}{0.0mm}
 \setlength{\parsep}{0.0mm}
 \setlength{\partopsep}{0.0mm}
 \setlength{\leftmargin}{15pt}
 \setlength{\labelsep}{5pt}
 \setlength{\labelwidth}{10pt}}}
{
 \end{list} 
}

\newenvironment{varitemizeii}
{
\begin{list}{\labelitemiv}
{\setlength{\itemsep}{0.0mm}
 \setlength{\topsep}{0.0mm}
 \setlength{\parindent}{0.0mm}
 \setlength{\parskip}{0.0mm}
 \setlength{\parsep}{0.0mm}
 \setlength{\partopsep}{0.0mm}
 \setlength{\leftmargin}{15pt}
 \setlength{\labelsep}{5pt}
 \setlength{\labelwidth}{10pt}}}
{
 \end{list} 
}

\newcounter{number}

\begin{document}

\title[On Constructor Rewrite Systems and the Lambda-Calculus]{On Constructor Rewrite Systems\\ and the Lambda-Calculus\rsuper*}
\author[U.~Dal Lago]{Ugo Dal Lago}
\author[S.~Martini]{Simone Martini}
\address{Universit\`a di Bologna, and INRIA Sophia Antipolis}
\email{\{dallago, martini\}@cs.unibo.it}

\keywords{lambda calculus, term rewriting, implicit computational
  complexity} 
\subjclass{F.4.1}  \titlecomment{{\lsuper*}This paper is an extended
  version of~\cite{DLMicalp}, appeared in the proceedings of ICALP
  2009.}

\begin{abstract}\noindent
We prove that orthogonal constructor term rewrite systems and $\lambda$-calculus with weak 
(i.e., no reduction is allowed under  the scope of a $\lambda$-abstraction)
call-by-value reduction can simulate each other with a linear overhead.
In particular, weak call-by-value beta-reduction can be simulated by an orthogonal 
constructor term rewrite system in the same number of reduction steps. 
Conversely, each reduction in a term rewrite system can be simulated by a constant number 
of beta-reduction steps. This is relevant to implicit computational complexity, because the number of beta steps 
to normal form is polynomially related to the actual cost (that is, as performed on a Turing 
machine) of normalization, under weak call-by-value reduction. Orthogonal constructor term rewrite systems 
and $\lambda$-calculus are thus both polynomially related to Turing machines, taking as notion 
of cost their natural parameters.
\end{abstract}

\maketitle
\section*{Introduction}
\par
Implicit computational complexity is a young research area, whose main aim is the description of complexity phenomena 
based on language restrictions, and not on external 
measure conditions or on explicit machine models. 
It borrows techniques and results from mathematical logic (model theory, recursion theory, and proof theory)
and in doing so it has allowed the
incorporation of aspects of computational complexity into areas such as formal methods in software development
and programming language design. The most developed area of implicit computational complexity is probably the
model theoretic one --- finite model theory being a very successful way to describe
complexity classes. In the design of  programming language tools (e.g., type systems), however, syntactical
techniques prove more useful. In the last years we have seen much work restricting
recursive schemata and developing general proof theoretical techniques to enforce
resource bounds on programs. 
Important achievements have been the characterizations of several complexity classes
by means of limitations of recursive definitions (e.g.,~\cite{Bellantoni92CC,Leivant95RRI}) and, more recently, by 
using the ``\emph{light}'' fragments of 
linear logic~\cite{Girard98ic}. 
Moreover, rewriting techniques such as recursive path orderings and
the interpretation method have been proved useful in the field~\cite{Marion00}.
By borrowing the terminology from software design technology, we may dub this 
area as implicit computational complexity \emph{in the large}, aiming at a broad, global view on complexity classes.
We may have also an implicit computational complexity \emph{in the small} ---
using logic to study single machine-free models of computation. Indeed, many models of computations do not come
with a natural cost model --- a definition of cost which is both intrinsically rooted in the model of 
computation, and, at the same time, it is polynomially related to the cost of implementing that
model of computation on a standard Turing machine. The main example is the $\lambda$-calculus: the most natural intrinsic
parameter of a computation is its number of beta-reductions, but this very parameter bears no
relation, in general, with the actual cost of performing that computation, since a beta-reduction may involve the duplication of arbitrarily big subterms\footnote{
In full beta-reduction, the size of the duplicated term is indeed arbitrary and does not depend on
the size of the original term the reduction started from. The situation is much different with weak 
reduction, as we will see.}.
What we call implicit computational complexity in the small, therefore, gives complexity significance to
notions and results for computation models where such natural cost measures do not exist, or are
not obvious. In particular, it looks for cost-explicit simulations between such computational models.

The present paper applies this viewpoint to the relation between $\lambda$-calculus and orthogonal constructor 
term rewrite systems (OCRSs in the following). We will prove that these two computational
models simulate each other with a linear overhead. That each OCRS
could be simulated by $\lambda$-terms and beta-reduction is
well known, in view of the availability, in $\lambda$-calculus, of
fixed-point operators, which may be used to solve the mutual recursion expressed by 
first-order rewrite rules. 
Here (Section~\ref{Sect:CTR2L})
we make explicit the complexity content of this simulation, by showing that
any first-order rewriting of $n$ steps can be simulated by $kn$ beta steps, where
$k$ depends on the specific rewrite system but \emph{not} on the size of the involved terms.
Crucial to this result is the encoding of constructor terms using Scott's schema for numerals~\cite{Wadsworth80}.
Indeed, Parigot~\cite{Parigot89CSL} (see also~\cite{ParigotRoziere93}) shows that in the pure $\lambda$-calculus
Church numerals do not admit a predecessor working in a constant number of beta steps.
Moreover, Splawski and Urzyczyn~\cite{SplawskiU99} show that it is unlikely that our encoding 
could work in the typed context of System $\mathsf{F}$.

Section~\ref{sect:CRS} studies the converse -- the simulation of (weak) $\lambda$-calculus reduction by means of
OCRSs. We give an encoding of $\lambda$-terms into a (first-order) constructor term rewrite 
system. We write   $\LambdatoTRS{\cdot}$ for the map returning a first-order term, given a $\lambda$-term; 
$\LambdatoTRS{M}$ is, in a sense, a complete defunctionalization of the $\lambda$-term $M$, 
where any $\lambda$-abstraction is represented by an atomic constructor. This is similar, 
although not technically the same, to the use of supercombinators (e.g.,~\cite{PJ87}).
We show that $\lambda$-reduction is simulated step by step by first-order rewriting 
(Theorem~\ref{theo:termreducible}).

As a consequence, taking the number of beta steps as a cost model for weak $\lambda$-calculus
is equivalent (up to a linear function) to taking the number of rewritings in OCRSs
systems. This is relevant to implicit computational complexity ``in the small'', because
the number of beta steps to normal form is polynomially related to the
actual cost (that is, as performed on a Turing machine) of normalization,
under weak call-by-value reduction. This has been established by 
Sands, Gustavsson, and Moran~\cite{Sands:Lambda02}, by a fine analysis of a $\lambda$-calculus implementation based on
a stack machine. OCRSs and $\lambda$-calculus
are thus both \emph{reasonable} machines (see the ``invariance thesis'' in~\cite{vanEmdeBoas90}),
taking as notion of cost their natural, intrinsic parameters.

As a byproduct, in Section~\ref{Sect:GraphRep} we sketch a different proof
of the cited result in~\cite{Sands:Lambda02}. Instead of using a stack machine, 
we show how we could implement
constructor term rewriting via term graph rewriting. In term graph rewriting we
avoid the explicit duplication and substitution inherent to
rewriting (and thus also to beta-reduction) and, moreover, we exploit the possible sharing of subterms. 
A more in-depth study of the complexity of 
(constructor) graph rewriting and its relations with (constructor) term rewriting can be found in another
paper by the authors~\cite{DLM09}.

In Section~\ref{sect:HeadReduction}, we show how to obtain the same results of the previous sections when 
call-by-name replaces call-by-value as the underlying strategy in the $\lambda$-calculus.

 
\section{Preliminaries}\label{sect:prelim}
The language we study is the pure untyped
$\lambda$-calculus endowed with weak (that is, we never reduce
under an abstraction)
call-by-value reduction. 
\begin{defi}
The following definitions are standard:
\begin{varitemize}
  \item
    \emph{Terms} are defined as follows:
    $$
    \lambdaone::=\varone\;|\;\lambda\varone.\lambdaone\;|\;\lambdaone\lambdaone ,
    $$
    where $\varone$ ranges a denumerable set $\Variables$.
    $\Lambdaterms$ denotes the set of all $\lambda$-terms.
    We assume the existence of a fixed, total, order on $\Variables$; this
    way $\FV{\lambdaone}$ will be a sequence (without repetitions) of variables, not a set. A term
    $\lambdaone$ is said to be \emph{closed} if $\FV{\lambdaone}=\varepsilon$,
    where $\varepsilon$ is the empty sequence.
  \item
    Values are defined as follows:
    $$
    \valueone::=\varone\;|\;\lambda\varone.\lambdaone .
    $$
  \item
    Weak call-by-value reduction is denoted by $\rewrlambdav$
    and is obtained by closing call-by-value reduction under
    any applicative context:
    $$
    \begin{array}{ccccccccc}
      \infer{(\lambda\varone.\lambdaone)\valueone\rewrlambdav\lambdaone\{\valueone/\varone\}}{}
      &&&&
      \infer{\lambdaone\lambdathree\rewrlambdav\lambdatwo\lambdathree}{\lambdaone\rewrlambdav\lambdatwo}
      &&&&
      \infer{\lambdathree\lambdaone\rewrlambdav\lambdathree\lambdatwo}{\lambdaone\rewrlambdav\lambdatwo}
    \end{array}
    $$
    Here $\lambdaone, \lambdatwo, \lambdathree$ range over terms, while $\valueone$ ranges over values.
  \item
    The length $\length{M}$ of $M$ is defined as follows, by induction
    on $M$: $\length{\varone}=1$, $\length{\lambda\varone.\lambdaone}=\length{\lambdaone}+1$ 
    and $\length{\lambdaone\lambdatwo}=\length{\lambdaone}+\length{\lambdatwo}+1$.
\end{varitemize}
\end{defi}

\noindent Weak call-by-value reduction enjoys many nice properties. In
particular, the one-step diamond property holds and, as a consequence,
the number of beta steps to normal form (if any) is invariant on the
reduction order~\cite{CIE2006} (this justifies the way we defined
reduction, which is slightly more general than Plotkin's
one~\cite{Plotkin75tcs}). It is then meaningful to define
$\Time{\lambdaone}$ as \emph{the} number of beta steps to normal form
(or $\omega$ if such a normal form does not exist). This cost model
will be referred to as the \emph{unitary} cost model, since each beta
(weak call-by-value) reduction step counts for $1$ in the global cost
of normalization. Moreover, notice that $\alpha$-conversion is not
needed during reduction of closed terms: if
$\lambdaone\rewrlambdav\lambdatwo$ and $\lambdaone$ is closed, then
the reduced redex will be in the form
$(\lambda\varone.\lambdathree)\valueone$, where $\valueone$ is a
\emph{closed} value. As a consequence, arguments are always closed and
open variables cannot be captured. Suppose $\lambdaone$ has $n$ free
variables $\varone_1\leq\ldots\leq\varone_n$, and that
$\lambdatwo_1,\ldots,\lambdatwo_n$ are lambda-terms. The term
$\lambdaone\{\lambdatwo_1/\varone_1,\ldots,\lambdatwo_n/\varone_n\}$
is sometimes denoted simply with
$\lambdaone(\lambdatwo_1,\ldots,\lambdatwo_n)$, taking advantage of
the implicit order between the variables.

The following lemma gives us a generalization of the fixed-point (call-by-value)
combinator (but observe the explicit limit $k$ on the reduction length, 
in the spirit of implicit computational complexity in the small):
\begin{lem}[call-by-value fixpoint combinator]\label{lemma:mfpc}
For every natural number $n$, there are terms $H_1,\ldots,H_n$ and a natural number $m$ such
that for any sequence of values $V_1,\ldots,V_n$ and for any $1\leq i\leq n$:
$$
H_iV_1\ldots V_n\rewrlambdav^k V_i(\lambda x.H_1V_1\ldots V_nx)\ldots(\lambda x.H_nV_1\ldots V_nx),
$$
where $k\leq m$.
\end{lem}
\proof
The terms we are looking for are simply the following:
$$
H_i\equiv\lambdaone_i\lambdaone_1\ldots\lambdaone_n
$$
where, for every $1\leq j\leq n$,
$$
\lambdaone_j\equiv\lambda\varone_1.\ldots.\lambda\varone_n.\lambda\vartwo_1.\ldots.\vartwo_n.\vartwo_j
(\lambda\varthree.\varone_1\varone_1\ldots\varone_n\vartwo_1\ldots\vartwo_n\varthree)
\ldots
(\lambda\varthree.\varone_n\varone_1\ldots\varone_n\vartwo_1\ldots\vartwo_n\varthree).
$$
The natural number $m$ is simply $2n$.
\qed

We only consider orthogonal and constructor rewriting in this paper.
A \emph{constructor term rewrite system} is a pair 
$\TRSone=(\Functions{\TRSone},\Rules{\TRSone})$ where:
\begin{varitemize}
\item
  Symbols in the signature $\Functions{\TRSone}$ can be either
  \emph{constructors} or \emph{function symbols}, each with its arity.
  \begin{varitemizeii} 
    \item
      Terms in $\TRScontermsp{\TRSone}$ are those built
      from constructors and are called \emph{constructor terms}.
    \item
      Terms in $\TRSpatsp{\TRSone}$ are those built
      from constructors and variables and are called \emph{patterns}.
    \item
      Terms in $\TRStermsp{\TRSone}$ are those built
      from constructor and function symbols and are called \emph{closed terms}.
    \item
      Terms in $\TRSvartermsp{\TRSone}$ are those built
      from constructors, functions symbols and variables in $\Variables$ and are dubbed
      \emph{terms}.
    \end{varitemizeii}
\item
  Rules in $\Rules{\TRSone}$ are in the form $\funone(\patone_1,\ldots,\patone_n)\rewr{\TRSone}\termone$
  where $\funone$ is a function symbol, $\patone_1,\ldots,\patone_n\in\TRSpatsp{\TRSone}$
  and $t\in\TRSvartermsp{\TRSone}$.
  We here consider orthogonal rewrite systems only, i.e. we assume that no distinct two
  rules in $\Rules{\TRSone}$ are overlapping and that every variable appears at most
  once in the lhs of any rule in $\Rules{\TRSone}$. Moreover, we assume that reduction is
  call-by-value, i.e. the substitution triggering any reduction must assign
  \emph{constructor} terms to variables. This restriction is anyway natural in
  constructor rewriting.
\end{varitemize}
For any term $\termone$ in a OCRS, $\length{\termone}$ denotes 
the number of symbol occurrences, while $\plength{\termone}{\funone}$ denotes
the number of occurrences of the symbol $\funone$ in $\termone$.
Similarly to $\lambda$-terms, if $\termone$ contains instances of $n$ variables
$\varone_1\leq\ldots\leq\varone_n$, the term 
$\termone\{\termtwo_1/\varone_1,\ldots,\termtwo_n/\varone_n\}$
is sometimes denoted simply with $\termone(\termtwo_1,\ldots,\termtwo_n)$.
\section{From $\lambda$-Calculus to Constructor Term Rewriting}\label{sect:CRS}
In this section, we will prove that the $\lambda$-calculus, in the form introduced 
in Section~\ref{sect:prelim}, can be seen as a OCRS. This result will be spelled
out as follows.
\begin{varitemize}
\item
  An OCRS $\TRS$ on a signature $\Functions{\TRS}$ will be defined, together with two maps
  $\LambdatoTRS{\cdot}:\Lambdaterms\rightarrow\TRSvartermsp{\Functions{\TRS}}$ and
  $\TRStolambda{\cdot}:\TRSvartermsp{\Functions{\TRS}}\rightarrow\Lambdaterms$.
  These two maps are \emph{not} bijections. However, $\LambdatoTRS{\cdot}$
  is injective, and $\TRStolambda{\cdot}\circ\LambdatoTRS{\cdot}$ is the identity.
\item
  The concept of canonicity for terms in $\TRSvartermsp{\TRS}$ will be defined. Moreover,
  the set of canonical terms will be shown to include $\LambdatoTRS{\Lambdaterms}$
  and to be closed by reduction.
\item
  Reduction of canonical terms will be shown to simulate weak call-by-value reduction
  on $\lambda$-terms, via $\TRStolambda{\cdot}$. Conversely, the dynamics of
  $\lambda$-terms is proved to simulate rewriting of constructor terms again through
  $\TRStolambda{\cdot}$.
\end{varitemize}
Altogether, the three ingredients above implies that $\TRS$ is a sound and complete
way of implementing call-by-value $\beta$-reduction.

Let us start by defining $\TRS$ and the two functions allowing to translate terms
$\TRS$ into $\lambda$-terms and, conversely, $\lambda$-terms back into terms of $\TRS$. 
Canonicity can already be defined.
\begin{defi}[The OCRS $\TRS$, Canonicity]
The OCRS $\TRS$ is defined as a set of rules $\Rules{\TRS}$ over
an infinite signature $\Functions{\TRS}$. In particular:
\begin{varitemize} 
  \item
    The signature $\Functions{\TRS}$ includes the binary function symbol $\appTRS$ and
    constructor symbols $\constr{\varone}{M}$ for every $\lambdaone\in\Lambdaterms$ and
    every $\varone\in\Variables$. The arity of
    $\constr{\varone}{\lambdaone}$ is the length of $\FV{\lambda\varone.\lambdaone}$. 
    To every term $\lambdaone\in\Lambdaterms$ we can associate
    a term $\LambdatoTRS{\lambdaone}\in\TRSvarterms$ as follows:
    \begin{eqnarray*}
      \LambdatoTRS{\varone}&=&\varone;\\
      \LambdatoTRS{\lambda\varone.\lambdaone}&=&\constr{\varone}{\lambdaone}(\varone_1,\ldots,\varone_{n}),
        \mbox{ where $\FV{\lambda\varone.\lambdaone}=\varone_1,\ldots,\varone_n$};\\
      \LambdatoTRS{\lambdaone\lambdatwo}&=&\appTRS(\LambdatoTRS{\lambdaone},\LambdatoTRS{\lambdatwo}).
    \end{eqnarray*}
    Observe that if $\lambdaone$ is closed, then $\LambdatoTRS{\lambdaone}\in\TRSterms$. 
  \item
    The rewrite rules in $\Rules{\TRS}$ are all the rules in the following form:
    $$
    \appTRS(\constr{\varone}{\lambdaone}(\varone_1,\ldots,\varone_{n}),x)\rewrTRS\LambdatoTRS{M},
    $$
    where $\FV{\lambda\varone.\lambdaone}=\varone_1,\ldots,\varone_n$.
  \item
    To every term $\termone\in\TRSvarterms$ we can associate
    a term $\TRStolambda{\termone}\in\Lambdaterms$ as follows:
    \begin{eqnarray*}
      \TRStolambda{\varone}&=&\varone\\
      \TRStolambda{\appTRS(\termtwo,\termthree)}&=&\TRStolambda{\termtwo}\TRStolambda{\termthree}\\
      \TRStolambda{\constr{\varone}{\lambdaone}(\termone_1,\ldots\termone_n)}&=&
      (\lambda\varone.\lambdaone)\{\TRStolambda{\termone_1}/\varone_1,\ldots,\TRStolambda{\termone_n}/\varone_n\}
    \end{eqnarray*}
    where $\FV{\lambda\varone.\lambdaone}=\varone_1,\ldots,\varone_n$.
  \item
    A term $\termone\in\TRSterms$ is \emph{canonical} if either $\termone\in\TRSconterms$ or
    $\termone=\appTRS(\termtwo,\termthree)$ where $\termtwo$ and $\termthree$ are
    themselves canonical.
  \end{varitemize}
\end{defi}

\noindent Notice that the signature $\Functions{\TRS}$ contains an
infinite number of constructors.
\begin{exa}
Consider the $\lambda$-term $\lambdaone=(\lambda \varone.\varone\varone)(\lambda \vartwo.\vartwo\vartwo)$.
$\LambdatoTRS{\lambdaone}$ is
$\termone\equiv\appTRS(\constr{\varone}{\varone\varone},\constr{\vartwo}{\vartwo\vartwo})$.
Moreover, $\termone\rewrTRS\appTRS(\constr{\vartwo}{\vartwo\vartwo},\constr{\vartwo}{\vartwo\vartwo})\equiv\termtwo$, 
as expected. We have $\termtwo\rewrTRS\termtwo$. Both $\termone$ and $\termtwo$ are
canonical. Finally, $\TRStolambda{\termtwo}=(\lambda\vartwo.\vartwo\vartwo)(\lambda\vartwo.\vartwo\vartwo)$.
\end{exa}
The map $\LambdatoTRS{\cdot}$ is injective, but not surjective. However:
\begin{lem}\label{lemma:invert}
For every $\lambda$-term $\lambdaone\in\Lambdaterms$, $\TRStolambda{\LambdatoTRS{\lambdaone}}=\lambdaone$.
\end{lem}
\proof
By induction on $\lambdaone$:
\begin{varitemize}
\item
  If $\lambdaone=\varone$, then 
  $$
    \TRStolambda{\LambdatoTRS{\lambdaone}}=\TRStolambda{\LambdatoTRS{\varone}}=\TRStolambda{\varone}=\varone.
  $$
\item
  If $\lambdaone=\lambdatwo\lambdathree$, then
  $$
  \TRStolambda{\LambdatoTRS{\lambdaone}}=
  \TRStolambda{\appTRS(\LambdatoTRS{\lambdatwo},\LambdatoTRS{\lambdathree})}=
  \TRStolambda{\LambdatoTRS{\lambdatwo}}\TRStolambda{\LambdatoTRS{\lambdathree}}=
  \lambdatwo\lambdathree.
  $$
\item
  If $\lambdaone=\lambda\vartwo.\lambdatwo$, then
  $$
    \TRStolambda{\LambdatoTRS{\lambdaone}}=
    \TRStolambda{\constr{\vartwo}{\lambdatwo}(\varone_{1},\ldots,\varone_{n})}=
    (\lambda\vartwo.\lambdatwo)\{\varone_1/\varone_1,\ldots,\varone_n/\varone_n\}=
    \lambda\vartwo.\lambdatwo=\lambdaone.
  $$
\end{varitemize}
This concludes the proof.
\qed
Canonicity holds for terms in $\TRS$ obtained as images of (closed) 
$\lambda$-terms via $\LambdatoTRS{\cdot}$. Moreover, canonicity
is preserved by reduction in $\TRS$:
\begin{lem}\label{lemma:canonicity}
For every closed $\lambdaone\in\Lambdaterms$, $\LambdatoTRS{\lambdaone}$ is canonical.
Moreover, if $\termone$ is canonical and $\termone\rewrTRS\termtwo$, 
then $\termtwo$ is canonical. 
\end{lem}
\proof
$\LambdatoTRS{\lambdaone}$ is canonical for any $\lambdaone\in\Lambdaterms$ by
induction on the structure of $\lambdaone$ (which, by hypothesis, is either an
abstraction or an application $\lambdatwo\lambdathree$ where both
$\lambdatwo$ and $\lambdathree$ are closed). We can further prove that
$\termthree=\LambdatoTRS{\lambdaone}\{\termone_1/\varone_1,\ldots\termone_n/\varone_n\}$
is canonical whenever $\termone_1,\ldots,\termone_n\in\TRSconterms$ and
$\varone_1,\ldots,\varone_n$ includes all the variables in $\FV{\lambdaone}$:
\begin{varitemize}
\item
  If $\lambdaone=\varone_i$, then $\termthree=\termone_i$, which is
  clearly canonical.
\item
  If $\lambdaone=\lambdatwo\lambdathree$, then
  \begin{eqnarray*}
  \termthree&=&\LambdatoTRS{\lambdatwo\lambdathree}\{\termone_1/\varone_1,\ldots\termone_n/\varone_n\}\\
     &=&\appTRS\left(\LambdatoTRS{\lambdatwo}\{\termone_1/\varone_1,\ldots\termone_n/\varone_n\},
        \LambdatoTRS{\lambdathree}\{\termone_1/\varone_1,\ldots\termone_n/\varone_n\}\right)
  \end{eqnarray*}
  which is canonical, by IH.
\item
  If $\lambdaone=\lambda\vartwo.\lambdatwo$, then
  \begin{eqnarray*}
  \termthree&=&\LambdatoTRS{\lambda\vartwo.\lambdatwo}\{\termone_1/\varone_1,\ldots\termone_n/\varone_n\}\\
     &=&\constr{\vartwo}{\lambdatwo}(\varone_{i_1},\ldots,\varone_{i_m})\{\termone_1/\varone_1,\ldots\termone_n/\varone_n\}\\
     &=&\constr{\vartwo}{\lambdatwo}(\termone_{i_1},\ldots,\termone_{i_m})
  \end{eqnarray*}
  which is canonical, because each $\termone_i$ (and hence also $\termthree$) is in $\TRSconterms$.
\end{varitemize}
This implies the rhs of any instance of a rule in $\Rules{\TRS}$ is canonical. As a consequence,
$\termtwo$ is canonical whenever $\termone\rewrTRS\termtwo$ and $\termone$ is canonical.
This concludes the proof. 
\qed
For canonical terms, being a normal form is equivalent to being mapped to a normal
form via $\TRStolambda{\cdot}$. This is not true, in general: take as a counterexample
$\constr{\varone}{\vartwo}(\appTRS(\constr{\varthree}{\varthree},\constr{\varthree}{\varthree}))$,
which corresponds to $\lambda\varone.(\lambda\varthree.\varthree)(\lambda\varthree.\varthree)$ via $\TRStolambda{\cdot}$.
\begin{lem}\label{lemma:NFcanonical}
A canonical term $\termone$ is a normal form iff 
$\TRStolambda{\termone}$ is a normal form.
\end{lem}
\proof
If a canonical $\termone$ is a normal form, then $\termone$ does not contain the function symbol $\appTRS$ and,
as a consequence, $\TRStolambda{\termone}$ is an abstraction, which is always a normal form.
Conversely, if $\TRStolambda{\termone}$ is a normal form, then
$\termone$ is not in the form $\appTRS(\termtwo,\termthree)$, because otherwise
$\TRStolambda{\termone}$ will be a (closed) application, which cannot be a normal form.
But since $\termone$ is canonical, $\termone\in\TRSconterms$, which only contains terms
in normal form.
\qed
The following substitution lemma will be useful later.
\begin{lem}[Substitution]
For every term $\termone\in\TRSvarterms$ and every $\termone_1,\ldots,\termone_n\in\TRSconterms$,
$$
\TRStolambda{\termone\{\termone_1/\varone_1,\ldots,\termone_n/\varone_n\}}=
\TRStolambda{\termone}\{\TRStolambda{\termone_1}/\varone_1,\ldots,\TRStolambda{\termone_n}/\varone_n\}
$$
whenever $\varone_1,\ldots,\varone_n$ includes all the variables in $\termone$.
\end{lem}
\proof
By induction on $\termone$:
\begin{varitemize}
  \item
    If $\termone=\varone_i$, then 
    \begin{eqnarray*}
      \TRStolambda{\termone\{\termone_1/\varone_1,\ldots,\termone_n/\varone_n\}}&=&
      \TRStolambda{\varone_i\{\termone_1/\varone_1,\ldots,\termone_n/\varone_n\}}\\
      &=&\TRStolambda{\termone_i}\\
      &=&\varone_i\{\TRStolambda{\termone_1}/\varone_1,\ldots,\TRStolambda{\termone_n}/\varone_n\}\\
      &=&\termone\{\TRStolambda{\termone_1}/\varone_1,\ldots,\TRStolambda{\termone_n}/\varone_n\}.
    \end{eqnarray*}
  \item
    If $\termone=\appTRS(\termtwo,\termthree)$, then
    \begin{eqnarray*}
      \TRStolambda{\termone\{\termone_1/\varone_1,\ldots,\termone_n/\varone_n\}}&=&
      \TRStolambda{\appTRS(\termtwo,\termthree)\{\termone_1/\varone_1,\ldots,\termone_n/\varone_n\}}\\
      &=&\TRStolambda{\appTRS(\termtwo\{\termone_1/\varone_1,\ldots,\termone_n/\varone_n\},
        \termthree\{\termone_1/\varone_1,\ldots,\termone_n/\varone_n\})}\\
      &=&\TRStolambda{\termtwo\{\termone_1/\varone_1,\ldots,\termone_n/\varone_n\}}
      \TRStolambda{\termthree\{\termone_1/\varone_1,\ldots,\termone_n/\varone_n\}}\\
      &=&\TRStolambda{\termtwo}\{\TRStolambda{\termone_1}/\varone_1,\ldots,\TRStolambda{\termone_n}/\varone_n\}
      \TRStolambda{\termthree}\{\TRStolambda{\termone_1}/\varone_1,\ldots,\TRStolambda{\termone_n}/\varone_n\}\\
      &=&\TRStolambda{\termtwo}\TRStolambda{\termthree}
      \{\TRStolambda{\termone_1}/\varone_1,\ldots,\TRStolambda{\termone_n}/\varone_n\}\\
      &=&\TRStolambda{\appTRS(\termtwo,\termthree)}\{\TRStolambda{\termone_1}/\varone_1,\ldots,\TRStolambda{\termone_n}/\varone_n\}\\
      &=&\TRStolambda{\termone}\{\TRStolambda{\termone_1}/\varone_1,\ldots,\TRStolambda{\termone_n}/\varone_n\}.
    \end{eqnarray*}
  \item
    If $\termone=\constr{\vartwo}{\lambdatwo}(\termtwo_{1},\ldots,\termtwo_{m})$, then
    \begin{eqnarray*}
      \TRStolambda{\termone\{\termone_1/\varone_1,\ldots,\termone_n/\varone_n\}}&=&
      \TRStolambda{\constr{\vartwo}{\lambdatwo}(\termtwo_{1},\ldots,\termtwo_{m})
        \{\termone_1/\varone_1,\ldots,\termone_n/\varone_n\}}\\
      &=&\TRStolambda{\constr{\vartwo}{\lambdatwo}
        (\termtwo_{1}\{\termone_1/\varone_1,\ldots,\termone_n/\varone_n\},\ldots,
         \termtwo_{m}\{\termone_1/\varone_1,\ldots,\termone_n/\varone_n\})}\\
      &=&(\lambda\vartwo.\lambdatwo)\{
         \TRStolambda{\termtwo_{1}\{\termone_1/\varone_1,\ldots,\termone_n/\varone_n\}}/\varone_{i_1}\\
         & &\hspace{36pt},\ldots,\\
         & &\hspace{36pt}\TRStolambda{\termtwo_{m}\{\termone_1/\varone_1,\ldots,\termone_n/\varone_n\}}/\varone_{i_m}\}\\
      &=&(\lambda\vartwo.\lambdatwo)\{\TRStolambda{\termtwo_{1}}\{\TRStolambda{\termone_1}/\varone_1,\ldots,
         \TRStolambda{\termone_n}/\varone_n\}/\varone_{i_1}\\
         & &\hspace{36pt},\ldots,\\
         & &\hspace{36pt}\TRStolambda{\termtwo_m}\{\TRStolambda{\termone_1}/\varone_1,\ldots,\TRStolambda{\termone_n}/\varone_n\}/\varone_{i_m}\}\\
      &=&((\lambda\vartwo.\lambdatwo)\{\TRStolambda{\termtwo_{1}}/\varone_1,\ldots,\termtwo_{_m}/\varone_{i_1}\})
         \{\TRStolambda{\termone_1}/\varone_1,\ldots,\TRStolambda{\termone_n}/\varone_n\}\\
      &=&\TRStolambda{\constr{\vartwo}{\lambdatwo}(\termtwo_{1},\ldots,\termtwo_{m})}\{\TRStolambda{\termone_1}/\varone_1,\ldots,\TRStolambda{\termone_n}/\varone_n\}\\
      &=&\TRStolambda{\termone}\{\TRStolambda{\termone_1}/\varone_1,\ldots,\TRStolambda{\termone_n}/\varone_n\}.
    \end{eqnarray*}
\end{varitemize}
This concludes the proof.
\qed
Two of the previous lemmas imply that if $\lambdaone\in\Lambdaterms$, 
$\termone_1,\ldots,\termone_n\in\TRSconterms$ and $\varone_1,\ldots,\varone_n$ 
includes all the variables in $\FV{\lambdaone}$, then:
\begin{equation}\label{equat:commute}
\TRStolambda{\LambdatoTRS{\lambdaone}\{\termone_1/\varone_1,\ldots,\termone_n/\varone_n\}}=
\lambdaone\{\TRStolambda{\termone_1}/\varone_1,\ldots,\TRStolambda{\termone_n}/\varone_n\}.
\end{equation}
Reduction in $\TRS$ can be simulated by reduction in the $\lambda$-calculus,
provided the starting term is canonical.  
\begin{lem}\label{lemma:TRStolam}
If $\termone$ is canonical and $\termone\rewrTRS\termtwo$, then
$\TRStolambda{\termone}\rewrlambdav\TRStolambda{\termtwo}$.
\end{lem}
\proof
Consider the (instance of the) rewrite rule which
turns $\termone$ into $\termtwo$. Let it be
$$
\appTRS(\constr{\vartwo}{\lambdaone}(\termone_1,\ldots,\termone_n),\termthree)\rewrTRS
\LambdatoTRS{\lambdaone}\{\termone_1/\varone_1,\ldots,\termone_n/\varone_n,\termthree/\vartwo\}.
$$
Clearly,
$$
\TRStolambda{\appTRS(\constr{\vartwo}{\lambdaone}(\termone_1,\ldots,\termone_n),\termthree)}=
  ((\lambda\vartwo.\lambdaone)\{\TRStolambda{\termone_1}/\varone_1,\ldots,\TRStolambda{\termone_n}/\varone_n\})\TRStolambda{\termthree}
$$
while, by~(\ref{equat:commute}):
$$
\TRStolambda{\LambdatoTRS{\lambdaone}\{\termone_1/\varone_1,\ldots,\termone_n/\varone_n,\termthree/\vartwo\}}
=\lambdaone\{\TRStolambda{\termone_1}/\varone_1,\ldots,\TRStolambda{\termone_n}/\varone_n,\TRStolambda{\termthree}/\vartwo\}
$$
which implies the thesis.
\qed
Conversely, call-by-value reduction in the $\lambda$-calculus can be simulated in $\TRS$:
\begin{lem}\label{lemma:lamtoTRS}
If $\lambdaone\rewrlambdav\lambdatwo$, $\termone$ is canonical and $\TRStolambda{\termone}=\lambdaone$, then
$\termone\rewrTRS\termtwo$, where $\TRStolambda{\termtwo}=\lambdatwo$.
\end{lem}
\proof
Let $(\lambda\varone.\lambdathree)\valueone$ be the redex fired in $\lambdaone$ when rewriting
it to $\lambdatwo$. There must be a corresponding subterm $\termthree$ of $\termone$ such
that $\TRStolambda{\termthree}=(\lambda\varone.\lambdathree)\valueone$. Then
$$
\termthree=\appTRS(\constr{\varone}{\lambdafour}(\termone_1,\ldots,\termone_n),\termfour),
$$
where $\TRStolambda{\constr{\varone}{\lambdafour}(\termone_1,\ldots,\termone_n)}=\lambda\varone.\lambdathree$.
and $\TRStolambda{\termfour}=\valueone$. Observe that, by definition,
$$
\TRStolambda{\constr{\varone}{\lambdafour}(\termone_1,\ldots,\termone_n)}=
(\lambda\varone.\lambdafour)\{\TRStolambda{\termone_1}/\varone_1,\ldots,\TRStolambda{\termone_n}/\varone_n\}
$$
where $\FV{\lambdafour}=\varone_1,\ldots,\varone_n$. Since $\termone$ is canonical, 
$\termone_1,\ldots,\termone_n\in\TRSconterms$. Moreover, since $\valueone$ is a value,
$\termfour$ itself is in $\TRSconterms$. This implies
$$
\appTRS(\constr{\varone}{\lambdafour}(\termone_1,\ldots,\termone_n),\termfour)\rewrTRS
\LambdatoTRS{\lambdafour}\{\termone_1/\varone_1,\ldots,\termone_n/\varone_n,\termfour/\varone\}.
$$   
By~(\ref{equat:commute}):
\begin{eqnarray*}
\TRStolambda{\LambdatoTRS{\lambdafour}\{\termone_1/\varone_1,\ldots,\termone_n/\varone_n,\termfour/\varone\}}&=&
   \lambdafour\{\TRStolambda{\termone_1}/\varone_1,\ldots,\TRStolambda{\termone_n}/\varone_n,\TRStolambda{\termfour}/\varone\}\\
&=&(\lambdafour\{\TRStolambda{\termone_1}/\varone_1,\ldots,\TRStolambda{\termone_n}/\varone_n\})\{\TRStolambda{\termfour}/\varone\}\\
&=&(\lambda\varone.\lambdathree)\{\valueone/\varone\}.
\end{eqnarray*}
This concludes the proof.
\qed
The previous lemmas together imply the following theorem, by which $\lambda$-calculus
normalization can be mimicked (step-by-step) by reduction in $\TRS$:
\begin{thm}[Term Reducibility]\label{theo:termreducible}
Let $\lambdaone\in\Lambdaterms$ be a closed term. The following
two conditions are equivalent:
\begin{enumerate}[\em 1.]
\item
  $\lambdaone\rewrlambdav^n\lambdatwo$ where $\lambdatwo$ is in normal form;
\item
  $\LambdatoTRS{\lambdaone}\rewrTRS^n\termone$ where
  $\TRStolambda{\termone}=\lambdatwo$ and $\termone$ is in normal form.
\end{enumerate}
\end{thm}
\proof
Suppose $\lambdaone\rewrlambdav^n\lambdatwo$, where $\lambdatwo$ is in normal form. 
Then, by applying Lemma~\ref{lemma:lamtoTRS}, we obtain a term $\termone$ such that
$\LambdatoTRS{\lambdaone}\rewrTRS^n\termone$ and $\TRStolambda{\termone}=\lambdatwo$.
By Lemma~\ref{lemma:canonicity}, $\termone$ is canonical and, by Lemma~\ref{lemma:NFcanonical}, 
it is in normal form. Now, suppose $\LambdatoTRS{\lambdaone}\rewrTRS^n\termone$ where
$\TRStolambda{\termone}=\lambdatwo$ and $\termone$ is in normal form. By
applying $n$ times Lemma~\ref{lemma:TRStolam}, we obtain
$\TRStolambda{\LambdatoTRS{\lambdaone}}\rewrlambdav^n\TRStolambda{\termone}=\lambdatwo$.
But $\TRStolambda{\LambdatoTRS{\lambdaone}}=\lambdaone$ by Lemma~\ref{lemma:invert}
and $\lambdatwo$ is a normal form by Lemma~\ref{lemma:NFcanonical}, since $\LambdatoTRS{\lambdaone}$ and $\termone$ are
canonical by Lemma~\ref{lemma:canonicity}.
\qed
There is another nice property of $\TRS$, that will be crucial in proving the main
result of this paper:
\begin{proposition}[Subterm Property]\label{prop:constred}
For every $\lambdaone\in\Lambdaterms$, for every $\termone$ with $\LambdatoTRS{\lambdaone}\rewrTRS^*\termone$
and for every occurrence of a constructor $\constr{\varone}{\lambdatwo}$ in $\termone$, $\lambdatwo$ is
a subterm of $\lambdaone$.
\end{proposition}
\proof
Assume $\LambdatoTRS{\lambdaone}\rewrTRS^n\termone$ and proceed
by induction on $n$.
\qed
\begin{exa}
Let us consider the $\lambda$-term $\lambdaone=(\lambda\varone.(\lambda \vartwo.\varone)\varone)
(\lambda\varthree.\varthree)$. Notice that 
$$
\lambdaone\rewrlambdav(\lambda \vartwo.(\lambda\varthree.\varthree))
(\lambda\varthree.\varthree)\rewrlambdav\lambda\varthree.\varthree.
$$
Clearly $\LambdatoTRS{\lambdaone}=\appTRS(\constr{\varone}{(\lambda\vartwo.\varone)\varone},
\constr{\varthree}{\varthree})$. Moreover:
$$
\appTRS(\constr{\varone}{(\lambda\vartwo.\varone)\varone},
\constr{\varthree}{\varthree})\rewrTRS \appTRS(\constr{\vartwo}{\varone}(\constr{\varthree}{\varthree}),
   \constr{\varthree}{\varthree})\rewrTRS\constr{\varthree}{\varthree}.
$$
For every constructor $\constr{\varfour}{\lambdatwo}$ occurring in any term in the previous
reduction sequence, $\lambdatwo$ is a subterm of $\lambdaone$. 
\end{exa}

A remark on $\TRS$ is now in order. $\TRS$ is an infinite OCRS, since $\Functions{\TRS}$ contains
an infinite amount of constructor symbols and, moreover, there are infinitely many
rules in $\Rules{\TRS}$. As a consequence, what we have presented here is an embedding
of the (weak, call-by-value) $\lambda$-calculus into an infinite OCRS. Consider, now,
the following scenario: suppose the $\lambda$-calculus is used to write a \emph{program} $\lambdaone$,
and suppose that inputs to $\lambdaone$ form an infinite set of $\lambda$-terms $\Theta$ which can 
anyway be represented by a finite set of constructors in $\TRS$. In this scenario,
Proposition~\ref{prop:constred} allows to conclude the existence of finite
subsets of $\Functions{\TRS}$ and $\Rules{\TRS}$ such that \emph{every} $\lambdaone\lambdatwo$
(where $\lambdatwo\in\Theta$) can be reduced via $\TRS$ by using only
constructors and rules in those \emph{finite} subsets. As a consequence, we can see the
above schema as one that puts any program $\lambdaone$ 
in correspondence to a \emph{finite} OCRS. Finally, observe that assuming \emph{data} to
be representable by a finite number of constructors in $\TRS$ is reasonable.
Scott's scheme~\cite{Wadsworth80}, for example, allows to represent any term in a given
free algebra in a finitary way, e.g. the natural number $0$ becomes
$\nat{0}\equiv\constr{\vartwo}{\lambda\varthree.\varthree}$ while
$n+1$ becomes $\nat{n+1}\equiv\constr{\vartwo}{\lambda\varthree.\vartwo\varone}(\nat{n})$.
Church's scheme, on the other hand, does not have this property.
\subsection{An Example}
Consider the lambda terms $\lambdaone=\lambda\varone.\lambda\vartwo.\varone\vartwo\varone$
and $\lambdatwo=\lambda\varone.\lambda\vartwo.\vartwo\varone\vartwo$. It is easy to
verify that:
\begin{align*}
\lambdathree\equiv(\lambdaone\lambdatwo)\lambdaone&\rewrlambdav^2(\lambdatwo\lambdaone)\lambdatwo,\\
\lambdafour\equiv(\lambdatwo\lambdaone)\lambdatwo&\rewrlambdav^2(\lambdaone\lambdatwo)\lambdaone.
\end{align*}
Therefore, both $\lambdathree$ and $\lambdafour$ diverge. Now:
\begin{align*}
\LambdatoTRS{\lambdathree}&\equiv\appTRS(\appTRS(\constr{\varone}{\lambda\vartwo.\varone\vartwo\varone},
   \constr{\varone}{\lambda\vartwo.\vartwo\varone\vartwo}),\constr{\varone}{\lambda\vartwo.\varone\vartwo\varone})\\
   &\rewrTRS\termone\equiv\appTRS(\constr{\vartwo}{\varone\vartwo\varone}(\constr{\varone}{\lambda\vartwo.\vartwo\varone\vartwo}),
   \constr{\varone}{\lambda\vartwo.\varone\vartwo\varone})\\
   &\rewrTRS\appTRS(\appTRS(\constr{\varone}{\lambda\vartwo.\vartwo\varone\vartwo},
   \constr{\varone}{\lambda\vartwo.\varone\vartwo\varone}),\constr{\varone}{\lambda\vartwo.\vartwo\varone\vartwo})\\
   &\equiv\LambdatoTRS{\lambdafour}.
\end{align*}
Similarly, $\LambdatoTRS{\lambdafour}\rewrTRS^2\LambdatoTRS{\lambdathree}$. Observe
that along the computation we reach the term $\termone$, which is not the image of any $\lambda$-term. However, all constructor terms in the reduction are canonical and, moreover,
$\TRStolambda{\termone}$ is $(\lambda\vartwo.\lambdatwo\vartwo\lambdatwo)\lambdaone$, 
the lambda term found along the reduction from $\lambdathree$ to $\lambdafour$.
\section{From Constructor Term Rewriting to the $\lambda$-Calculus}\label{Sect:CTR2L}
In this section, we will show that one rewriting step of any constructor rewrite
system can be simulated by a fixed number of weak call-by-value beta-reductions. As 
an easy consequence, $\lambda$-calculus will be shown to efficiently simulate any
OCRS. During this section we will assume fixed an  OCRS $\TRSone$  over a finite signature $\Functions{\TRSone}$. Let
$\conone_1,\ldots,\conone_g$ be the constructors of $\TRSone$
and let $\funone_1,\ldots,\funone_h$ be the function symbols of
$\TRSone$. We will describe several constructions, which work independently of $\TRSone$ (they only depends on the arity of the symbols).
\begin{varitemize}
\item
  A map $\TRSonetolambdaI{\cdot}:\TRScontermsp{\TRSone}\rightarrow\Lambdaterms$ can
  be defined by recursion on the structure of the input. The map can be
  extended to constructors of $\TRSone$ (which are \emph{not} terms by themselves),
  in such a way that for every $\conone_i$, the lambda term $\TRSonetolambdaI{\conone_i}$
  ``computes'' $\TRSonetolambdaI{\conone_i(\termone_1\ldots\termone_{\arity{\conone_i}})}$
  when fed with $\TRSonetolambdaI{\termone_1}\ldots\TRSonetolambdaI{\termone_{\arity{\conone_i}}}$,
  for any $\termone_1,\ldots,\termone_{\arity{\conone_i}}\in\TRScontermsp{\TRSone}$. (See Definition~\ref{Def-TRSonetolambdaI}.)
\item
  Defining a map analogous to $\TRSonetolambdaI{\cdot}$, but acting on closed terms (and not only on 
  constructor terms) is more delicate. Indeed, a term 
  $\funone_i(\termone_1\ldots\termone_{\arity{\conone_i}})$ 
  does not necessarily rewrite to a constructor term,
  even if it does \emph{not} diverge ---
  the rewrite rules of $\TRSone$ are not necessarily exhaustive and a deadlock
  can be reached. To handle this case we define a lambda term $\errorterm\in\Lambdaterms$, which 
  will represent any deadlocked term.
\item
  Now a map $\TRSonetolambdaII{\cdot}:\TRStermsp{\TRSone}\rightarrow\Lambdaterms$
  can be defined, in such a way that $\TRSonetolambdaII{\termone}$ reduces
  to $\TRSonetolambdaI{\termtwo}$ (where $\termtwo\in\TRScontermsp{\TRSone}$) if 
  $\termone$ has normal form $\termtwo$, but $\TRSonetolambdaII{\termone}$ reduces to $\errorterm$
  if $\termone$ rewrites to a deadlock. The map $\TRSonetolambdaII{\cdot}$
  is defined compositionally, that is to say:
  \begin{eqnarray*}
    \TRSonetolambdaII{\conone(\termone_1,\ldots,\termone_{\arity{\conone_i}})}&=&\TRSonetolambdaII{\conone_i}
    \TRSonetolambdaII{\termone_1}\ldots\TRSonetolambdaII{\termone_{\arity{\conone_i}}};\\
    \TRSonetolambdaII{\funone_i(\termone_1,\ldots,\termone_{\arity{\funone_i}})}&=&\TRSonetolambdaII{\funone_i}
    \TRSonetolambdaII{\termone_1}\ldots\TRSonetolambdaII{\termone_{\arity{\funone_i}}}.
  \end{eqnarray*}
  In other words, $\TRSonetolambdaII{\cdot}$ is completely specified by its
  behavior on constructors and function symbols.
\item
  While defining $\TRSonetolambdaII{\conone}$ is relatively easy 
  (Definition~\ref{Def-TRSonetolambdaII} and Lemma~\ref{Lemma-TRSonetolambdaII}),
  $\TRSonetolambdaII{\funone}$ requires a form of pattern matching
  to be implemented in the $\lambda$-calculus (Lemma~\ref{lemma:pm} and Definition~\ref{Def-TRSonetolambdaII-fun}).
\item The complete simulation is stated in Theorem~\ref{theo:simulcl}. The example in~\ref{sect-EsempioTRStoLam} may be used along the section to clarify the definitions.
\end{varitemize}
We will first concentrate on constructor terms, encoding them as
$\lambda$-terms using Scott's schema~\cite{Wadsworth80}. 
\begin{defi}\label{Def-TRSonetolambdaI}
\begin{varitemize}
\item
  Constructor terms can be easily put in correspondence with $\lambda$-terms
  by way of a map $\TRSonetolambdaI{\cdot}$ defined by induction as follows:
  $$
  \TRSonetolambdaI{\conone_i(\termone_1\ldots,\termone_n)}\equiv
  \lambda x_1.\ldots.\lambda x_g.\lambda y.x_i\TRSonetolambdaI{\termone_1}\ldots\TRSonetolambdaI{\termone_n}.
  $$
\item
  The function $\TRSonetolambdaI{\cdot}$ can be extended to a map on constructors:
  $$
  \TRSonetolambdaI{\conone_i}\equiv\lambda x_1.\ldots.\lambda x_{\arity{\conone_i}}.
  \lambda y_1.\ldots.\lambda y_g.\lambda z.y_ix_1\ldots x_{\arity{\conone_i}}.
  $$
  Trivially, if  $\termone_1,\ldots,\termone_n$ are in $\TRScontermsp{\TRS}$,
  $\TRSonetolambdaI{\conone_i}\TRSonetolambdaI{\termone_1}\ldots\TRSonetolambdaI{\termone_n}$
  rewrites to $\TRSonetolambdaI{\conone_i(\termone_1\ldots\termone_n)}$ in
  $\arity{\conone_i}$ steps. 
\item
  To represent an error value, we use the $\lambda$-term
  $\errorterm \equiv
  \lambda x_1.\ldots.\lambda x_g.\lambda y.y$. A $\lambda$-term which is either $\errorterm$ or in
  the form $\TRSonetolambdaI{\termone}$ is denoted with metavariables like
  $\cltermone$ or $\cltermtwo$.
\end{varitemize}
\end{defi}

\noindent The map $\TRSonetolambdaI{\cdot}$ defines encodings of constructor terms.
For function symbols our goal is defining another map $\TRSonetolambdaII{\cdot}$ returning
a $\lambda$-term given any term $t$ in $\TRStermsp{\TRSone}$, in such
a way that $\termone\rewrTRS^*\termtwo$ and $\termtwo\in\TRScontermsp{\TRSone}$ implies
$\TRSonetolambdaII{\termone}\rewrlambdav^*\TRSonetolambdaI{\termtwo}$. Moreover,
$\TRSonetolambdaII{\termone}$ should rewrite to $\errorterm$ whenever the rewriting of $\termone$
causes an error (i.e. $\termone$ has a normal form containing
a function symbol).
First of all, we define the $\lambda$-term $\TRSonetolambdaII{\conone_i}$ corresponding
to a constructor $\conone_i$.
\newcommand{\conslambda}{\mathit{CON}}
\newcommand{\patlambda}{\mathit{PAT}}
\begin{defi}\label{Def-TRSonetolambdaII}
  \begin{varitemize}
  \item
    For every $1\leq i\leq g$, for every $0\leq m\leq\arity{\conone_i}$, and for every
    sequence of variables $\varone_1,\ldots,\varone_m$, 
    define the $\lambda$-term $\conslambda^i_{\varone_1,\ldots,\varone_m}$ 
    by induction on $\arity{\conone_i}-m$:
    \begin{eqnarray*}
      \conslambda^i_{\varone_1,\ldots,\varone_{\arity{\conone_i}}}&\equiv&
      \lambda\vartwo_1.\ldots.\lambda\vartwo_{g}.\varthree.\vartwo_{i}\varone_1\ldots\varone_{\arity{\conone_i}};\\
      \forall m: 0\leq m<\arity{\conone_i} \qquad \conslambda^i_{\varone_1,\ldots,\varone_m}&\equiv&
      \lambda\vartwo.\vartwo\lambdatwo_{1,i}^m\ldots\lambdatwo_{g,i}^m\lambdathree^m_i;
    \end{eqnarray*}
    where:
    \begin{eqnarray*}
      \lambdatwo_{j,i}^m&\equiv&\lambda\varthree_1.\ldots.\lambda\varthree_{\arity{\conone_j}}.
      (\lambda\varone_{m+1}.\conslambda^i_{\varone_1,\ldots,\varone_{m+1}})
      \conslambda^{j}_{\varthree_1,\ldots,\varthree_{\arity{\conone_j}}};\\
      \lambdathree^m_i&\equiv&\lambda\varthree_{m+2}.\ldots.\lambda\varthree_{\arity{\conone_i}}.\errorterm.
    \end{eqnarray*}
  \item
    For every $1\leq i\leq g$, the $\lambda$-term $\TRSonetolambdaII{\conone_i}$ is $\conslambda^i_{\varepsilon}$.
  \end{varitemize}
\end{defi}
We need to prove that $\TRSonetolambdaII{\conone_i}$ does what it is supposed to do. We show something slightly stronger:
\begin{lem}\label{Lemma-TRSonetolambdaII}
There is a constant $n\in\N$ such that for any $i$, for any $m$, and for any
$\TRSonetolambdaI{\termone_1},\ldots,\TRSonetolambdaI{\termone_{\arity{\conone_i}}}$ in $\TRScontermsp{\TRSone}$:
$$
\conslambda^i_{\varone_1,\ldots,\varone_m}\{\TRSonetolambdaI{\termone_1}/\varone_1,\ldots,\TRSonetolambdaI{\termone_m}/\varone_m\}
\TRSonetolambdaI{\termone_{m+1}}\ldots\TRSonetolambdaI{\termone_{\arity{\conone_i}}}
\rewrTRS^k\TRSonetolambdaI{\conone_i(\termone_1\ldots\termone_{\arity{\conone_i}})}
$$
where $k\leq n$, and
$$
\conslambda^i_{\varone_1,\ldots,\varone_m}\{\TRSonetolambdaI{\termone_1}/\varone_1,\ldots,\TRSonetolambdaI{\termone_m}/\varone_m\}
\cltermone_{m+1}\ldots\cltermone_{\arity{\conone_i}}
\rewrTRS^l\errorterm
$$
where $l\leq n$, whenever $\cltermone_{j}$ is either $\TRSonetolambdaI{\termone_j}$ or $\errorterm$
but at least one among $\cltermone_{m+1}\ldots\cltermone_{\arity{\conone_i}}$ is $\errorterm$.
\end{lem}
\proof
We proceed by induction on $\arity{\conone_i}-m$:
\begin{varitemize}
\item
  If $m=\arity{\conone_i}$, then
  \begin{eqnarray*}
    &&\conslambda^i_{\varone_1,\ldots,\varone_{\arity{\conone_i}}}\{\TRSonetolambdaI{\termone_1}/\varone_1,\ldots,\TRSonetolambdaI{\termone_{\arity{\conone_i}}}/\varone_{\arity{\conone_i}}\}\\
    &\equiv&(\lambda\vartwo_1.\ldots.\lambda\vartwo_{g}\vartwo_{i}\varone_1\ldots\varone_{\arity{\conone_i}})
      \{\TRSonetolambdaI{\termone_1}/\varone_1,\ldots,\TRSonetolambdaI{\termone_{\arity{\conone_i}}}/\varone_{\arity{\conone_i}}\}\\
    &\equiv&\lambda\vartwo_1.\ldots.\lambda\vartwo_{g}.\vartwo_{i}\TRSonetolambdaI{\termone_1}\ldots\TRSonetolambdaI{\termone_{\arity{\conone_i}}}\\
    &\equiv&\TRSonetolambdaI{\conone_i(\termone_1,\ldots,\termone_{\arity{\conone_i}})}.
  \end{eqnarray*}
\item
  If $m<\arity{\conone_i}$, we
  use the following abbreviations:
  \begin{eqnarray*}
    \lambdafour_{j,i}^m&\equiv&\lambdatwo_{j,i}^m\{\TRSonetolambdaI{\termone_1}/\varone_1,\ldots,\TRSonetolambdaI{\termone_{m}}/\varone_{m}\};\\
    \lambdafive_{j}^m&\equiv&\lambdathree_{j}^m\{\TRSonetolambdaI{\termone_1}/\varone_1,\ldots,\TRSonetolambdaI{\termone_{m}}/\varone_{m}\}.
  \end{eqnarray*}
  Let's distinguish two cases:
  \begin{varitemizeii}
  \item
    If $\cltermone_{m+1}\equiv\errorterm$, then:
    \begin{eqnarray*}
      &&\conslambda^i_{\varone_1,\ldots,\varone_{m}}\{\TRSonetolambdaI{\termone_1}/\varone_1,\ldots,
      \TRSonetolambdaI{\termone_{m}}/\varone_{m}\}\cltermone_{m+1}\ldots\cltermone_{\arity{\conone_i}}\\
      &\rewrlambdav&(\errorterm\lambdafour_{1,i}^m\ldots\lambdafour_{g,i}^m\lambdafive^m_i)\cltermone_{m+2}\ldots\cltermone_{\arity{\conone_i}}\\
      &\rewrlambdav^*&\lambdafive_i^m\cltermone_{m+2}\ldots\cltermone_{\arity{\conone_i}}\\
      &\rewrlambdav^*&\errorterm
    \end{eqnarray*}
  \item
    Let $\cltermone_{m+1}$ be $\TRSonetolambdaI{\termone_{m+1}}$,
    where $\termone_{m+1}\equiv\conone_j(\termtwo_{1},\ldots,\termtwo_{\arity{\conone_j}}) $. 
    Then:
    \begin{eqnarray*}
      &&\conslambda^i_{\varone_1,\ldots,\varone_{m}}\{\TRSonetolambdaI{\termone_1}/\varone_1,\ldots,
        \TRSonetolambdaI{\termone_{m}}/\varone_{m}\}\cltermone_{m+1}\ldots\cltermone_{\arity{\conone_i}}\\
      &\rewrlambdav&(\TRSonetolambdaI{\conone_j(\termtwo_{1},\ldots,\termtwo_{\arity{\conone_j}})}
        \lambdafour_{1,i}^m\ldots\lambdafour_{g,i}^m\lambdafive^m_i)\cltermone_{m+2}\ldots\cltermone_{\arity{\conone_i}}\\
      &\rewrlambdav^*&\lambdafour_{j,i}^m\TRSonetolambdaI{\termtwo_{1}}\ldots\TRSonetolambdaI{\termtwo_{\arity{\conone_j}}}
        \cltermone_{m+2}\ldots\cltermone_{\arity{\conone_i}}\\
      &\rewrlambdav^*&(\lambda\varone_{m+1}.\conslambda^i_{\varone_1,\ldots,\varone_{m+1}}\{\TRSonetolambdaI{\termone_1}/\varone_1,\ldots,
        \TRSonetolambdaI{\termone_{m}}/\varone_m\})\\
      &&(\conslambda^{j}_{\varthree_1,\ldots,\varthree_{\arity{\conone_j}}}\{\TRSonetolambdaI{\termtwo_1}/\vartwo_1,\ldots,
        \TRSonetolambdaI{\termone_{\arity{\conone_j}}}/\vartwo_{\arity{\conone_j}}\})\cltermone_{m+2}\ldots\cltermone_{\arity{\conone_i}}\\
      &\rewrlambdav^*&(\lambda\varone_{m+1}.\conslambda^i_{\varone_1,\ldots,\varone_{m+1}}\{\TRSonetolambdaI{\termone_1}/\varone_1,\ldots,
        \TRSonetolambdaI{\termone_{m}}/\varone_m\})\\
      &&(\TRSonetolambdaI{\conone_j(\termtwo_1,\ldots,\termtwo_{\arity{\conone_j}})})
        \cltermone_{m+2}\ldots\cltermone_{\arity{\conone_i}}\\
     &\rewrlambdav^*&\conslambda^i_{\varone_1,\ldots,\varone_{m+1}}\{\TRSonetolambdaI{\termone_1}/\varone_1,\ldots,
        \TRSonetolambdaI{\termone_{m+1}}/\varone_{m+1}\}\cltermone_{m+2}\ldots\cltermone_{\arity{\conone_i}}
    \end{eqnarray*}
    and, by the inductive hypothesis, the last term in the reduction sequence reduces to the correct
    normal form. The existence of a natural number $n$ with the prescribed properties is clear
    by observing that none of the reductions above have a length which depends on the parameters
    $\TRSonetolambdaI{\termone_1},\ldots,\TRSonetolambdaI{\termone_{m}}$ and $\cltermone_{m+1}\ldots\cltermone_{\arity{\conone_i}}$.
  \end{varitemizeii}
\end{varitemize}
This concludes the proof.
\qed
Interpreting function symbols is more difficult, since we have to ``embed'' the reduction rules
into the $\lambda$-term interpreting the function symbol. To do that, we need a preliminary
result to encode pattern matching. More specifically, suppose $\seqpone_1,\ldots,\seqpone_n$ are
non-overlapping sequences of patterns of the same length $m$, i.e. that for every
sequence of constructor terms $\termone_1,\ldots,\termone_m$ there is at most one $i$ with
$1\leq i\leq m$ such that $\termone_1,\ldots,\termone_m$ unifies with the patterns
in $\seqpone_i$. Then, we need to build a $\lambda$-term 
$\patlambda_{\seqpone_1,\ldots,\seqpone_n}^m$ which, when fed with $m$ (encodings of) constructor
terms and $n$ values, perform pattern matching and select the ``right'' value, or returns
$\errorterm$ if none of $\seqpone_1,\ldots,\seqpone_n$ unifies with the constructor terms
in input. 
\begin{lem}[Pattern matching]\label{lemma:pm}
Let $\seqpone_1,\ldots,\seqpone_n$ be non-overlapping sequences of patterns of the same length $m$.
Then there are a term $\patlambda_{\seqpone_1,\ldots,\seqpone_n}^m$ and an integer $l$ such that
for every sequence of values $\valueone_1,\ldots,\valueone_n$,
if $\seqpone_i=\patone_1,\ldots,\patone_m$ then
\begin{equation*}
\begin{split}
\patlambda_{\seqpone_1,\ldots,\seqpone_n}^m &
  \TRSonetolambdaI{\patone_1(\termone_1^1,\ldots,\termone_1^{k_1})}
  \ldots
  \TRSonetolambdaI{\patone_m(\termone_m^1,\ldots,\termone_m^{k_m})}
   \valueone_1\ldots \valueone_n\\
&\rewrlambdav^k \; \;
\valueone_i\TRSonetolambdaI{\termone_1^1}\ldots\TRSonetolambdaI{\termone_1^{k_1}}
\ldots\TRSonetolambdaI{\termone_m^1}\ldots\TRSonetolambdaI{\termone_m^{k_m}},
\end{split}
\end{equation*}
where $k\leq l$, whenever the $\termone_i^j$ are constructor terms. Moreover,
$$
\patlambda_{\seqpone_1,\ldots,\seqpone_n}^m\cltermone_1,\ldots,\cltermone_mV_1\ldots V_n
\rewrlambdav^k \errorterm,
$$
where $k\leq l$,
whenever $\cltermone_1,\ldots,\cltermone_m$ do not unify with
any of the sequences $\seqpone_1,\ldots,\seqpone_n$ or any of 
the $\cltermone_1,\ldots,\cltermone_m$ is itself $\errorterm$.
\end{lem}
\proof
We go by induction on $a=\sum_{i=1}^n\dsize{\seqpone_i}$, where
$\dsize{\seqpone_i}$ is the number of constructors occurrences in patterns
inside $\seqpone_i$:
\begin{varitemize}
\item
  If $a=0$ and $n=0$, then we should always return $\errorterm$: 
  $$
  \patlambda_{\varepsilon}^m\equiv\lambda\varone_1.\ldots.\lambda\varone_m.\errorterm.
  $$
\item
  If $a=0$ and $n>0$, then $n=1$ and $\seqpone_1$ is simply a sequence of variables 
  $\varone_1,\ldots,\varone_m$, because the $\seqpone_i$ are assumed to be non-overlapping.
  Then $\patlambda_{\varone_1,\ldots,\varone_m}^m$ is a term defined by induction on $m$
  which returns $\errorterm$ only if one of its first $m$ arguments is $\errorterm$ and otherwise
  returns its $m+1$-th argument applied to its first $m$ arguments.
\item
  If $a\geq 1$, then there must be integers $i$ and $j$ with 
  $1\leq i\leq m$ and $1\leq j\leq n$ such that
  $$
  \seqpone_j=\patone_1,\ldots,\patone_{i-1},\conone_k(\pattwo_1,\ldots,\pattwo_{\arity{\conone_k}}),
  \patone_{i+1},\ldots,\patone_{m}
  $$
  for a constructor $\conone_k$ and for some patterns $\patone_p$ and some $\pattwo_q$.
  Now, for every $1\leq p\leq n$ 
  and for every $1\leq j\leq g$ 
  we define sequences of patterns $\seqptwo_p^j$ and values $\valuetwo_p^j$ as follows:
  \begin{varitemizeii}
  \item
    If 
    $$
    \seqpone_p=\patone_1,\ldots,\patone_{i-1},\conone_j(\patthree_1,\ldots,\patthree_{\arity{\conone_j}}),
    \patone_{i+1}\ldots\patone_m
    $$
    then $\seqptwo_p^j$ is defined to be the sequence
    $$
    \patone_1,\ldots,\patone_{i-1},\patthree_1,\ldots,\patthree_{\arity{\conone_j}},\patone_{i+1},\ldots,\patone_m.
    $$
    Moreover, $\valuetwo_p$ is simply the indentity $\lambda\varone.\varone$.
  \item
    If 
    $$
    \seqpone_p=\patone_1,\ldots,\patone_{i-1},\conone_s(\patthree_1,\ldots,\patthree_{\arity{\conone_s}}),
    \patone_{i+1}\ldots\patone_m
    $$
    where $s\neq j$ then $\seqptwo_p^j$ and $\valuetwo_p^j$ are both undefined.
  \item
    Finally, if
    $$
    \seqpone_p=\patone_1,\ldots,\patone_{i-1},\varone,\patone_{i+1}\ldots\patone_m
    $$
    then $\seqptwo_p^j$ is defined to be the sequence
    $$
    \patone_1,\ldots,\patone_{i-1},\varone_1,\ldots,\varone_{\arity{\conone_j}},\patone_{i+1},\ldots,\patone_m.
    $$
    and $\valuetwo_p^j$ is the following $\lambda$-term
    $$
    \lambda\varone.\lambda\vartwo_1.\ldots.\lambda\vartwo_t.\varone_1.\ldots.\lambda\varone_{\arity{\conone_k}}.
    \lambda\varthree_1.\ldots.\lambda\varthree_u.\varone\vartwo_1\ldots\vartwo_t
    (\TRSonetolambdaI{\conone_j}\varone_1\ldots\varone_{\arity{\conone_j}})\varthree_1\ldots\varthree_u
    $$
    where $t$ is the number of variables in $\patone_1,\ldots,\patone_{i-1}$ and $u$ is the
    number of variables in $\patone_{i+1},\ldots,\patone_{m}$.
  \end{varitemizeii}
  As a consequence, for every $1\leq j\leq g$, we can find a natural number $t_j$ and 
  a sequence of pairwise distinct natural numbers $i_1,\ldots,i_{t_j}$ such that 
  $\seqptwo^j_{i_1},\ldots,\seqptwo^j_{i_{t_j}}$ are exactly the sequences which can
  be defined by the above construction. We are now able to formally define
  $\patlambda_{\seqpone_1,\ldots,\seqpone_n}^m$; it is the term
  $$
  \lambda\varone_1.\ldots.\lambda\varone_m.\lambda\vartwo_1.\ldots.\lambda\vartwo_n.
  ((\varone_i\valuethree_1\ldots\valuethree_g\valuethree_\bot)\varone_1\ldots\varone_{i-1}\varone_{i+1}\ldots\varone_m)
  \vartwo_1\ldots\vartwo_n
  $$
  where
  \begin{eqnarray*}
    \forall 1\leq j\leq g.\valuethree_j&\equiv&\lambda\varthree_1.\ldots.\lambda\varthree_{\arity{\conone_j}}.
    \lambda\varone_1.\ldots.\lambda\varone_{i-1}.\lambda\varone_{i+1}.\ldots.\lambda\varone_m.
    \lambda\vartwo_1.\ldots.\lambda\vartwo_n.\\
    & &\patlambda_{\seqptwo^j_{i_1},\ldots,\seqptwo^j_{i_{t_j}}}^{m-1+\arity{\conone_j}}\varone_1\ldots\varone_{i-1}\varthree_1\ldots\varthree_{\arity{\conone_j}}
      \varone_{i+1}\ldots\varone_m(\valuetwo^j_{i_1}\vartwo_{i_1})\ldots(\valuetwo^j_{i_{t_j}}\vartwo_{i_{t_j}})\\
    \valuethree_\bot&\equiv&\lambda\varone_1.\ldots.\lambda\varone_{i-1}.\lambda\varone_{i+1}.\ldots.\lambda\varone_m.
    \lambda\vartwo_1.\ldots.\lambda\vartwo_n.\errorterm
  \end{eqnarray*}
  Notice that, for every $j$,
  $a>\sum_{v=1}^{t_j}\dsize{\seqptwo^j_v}$. Moreover, for every $j$ any
  $\seqptwo^j_v$ has the same length $m-1+\arity{\conone_j}$.
  This justifies the application of the
  induction hypothesis above. Informally, $\patlambda_{\seqpone_1,\ldots,\seqpone_n}^m$ first do some case
  analysis based on the shape of its $i$-th argument. Based on the topmost constructor in it,
  one between $\valuethree_1,\ldots,\valuethree_h,\valuethree_\bot$ is selected which itself
  do the rest of the pattern matching by way of $\patlambda_{\seqptwo^j_{i_1},\ldots,\seqptwo^j_{i_{t_j}}}^{m-1+\arity{\conone_j}}$.
\end{varitemize}
This concludes the proof.
\qed
Once a general form of pattern matching is available in the $\lambda$ calculus, we may define
the $\lambda$-term $\TRSonetolambdaII{\funone_i}$ interpreting the function symbol $\funone_i$.
\begin{defi}\label{Def-TRSonetolambdaII-fun}
For every function symbol $\funone_i$, let
$$
\funone_i(\seqpone_i^1)\rewrTRS\termone_i^1,\;\ldots\;,
\funone_i(\seqpone_i^{n_i})\rewrTRS\termone_i^{n_i}
$$
be the rules for $\funone_i$. Moreover, suppose that
the variables appearing in the patterns in $\seqpone_i^j$ are
$z_i^{j,1},\ldots,z_i^{j,m_{i,j}}$. Observe that the
sequences $\seqpone_i^1,\ldots,\seqpone_i^{n_i}$ all have
the same length $m$. Recall that we have a signature with function symbols $\funone_1,\ldots,\funone_h$. For any $1\leq i \leq h$ 
the $\lambda$-term $\TRSonetolambdaII{\funone_i}$
interpreting $\funone_i$ is defined to be:
$$
H_i\valueone_1\ldots\valueone_h
$$
where
\begin{eqnarray*}
\valueone_i&\equiv&\lambda x_1.\ldots.\lambda x_h.\lambda y_1.\ldots.\lambda y_{\arity{\funone_i}}.
\patlambda^m_{\seqpone_i^1,\ldots,\seqpone_i^n}y_1\ldots y_{\arity{\funone_i}}\valuetwo_i^1\ldots\valuetwo_i^{n_i};\\
\valuetwo_i^j&\equiv&\lambda\varthree_1.\ldots.\lambda\varthree_{m_{i,j}}.\TRSonetolambdaIII{\termone_i^j};
\end{eqnarray*}
whenever $1\leq i\leq h$ and $1\leq j\leq n_i$ and
$\TRSonetolambdaIII{\cdot}$ is defined by induction as follows:
\begin{eqnarray*}
\TRSonetolambdaIII{\varone}&=&\varone;\\
\TRSonetolambdaIII{\conone_i(\termone_1,\ldots,\termone_{\arity{\conone_i}})}&=&\TRSonetolambdaII{\conone_i}
  \TRSonetolambdaIII{\termone_1}\ldots\TRSonetolambdaIII{\termone_{\arity{\conone_i}}};\\
\TRSonetolambdaIII{\funone_i(\termone_1,\ldots,\termone_{\arity{\funone_i}})}&=&\varone_i
  \TRSonetolambdaIII{\termone_1}\ldots\TRSonetolambdaIII{\termone_{\arity{\funone_i}}}.
\end{eqnarray*}
\end{defi}
We have now implicitly defined how the map $\TRSonetolambdaII{\cdot}$ behaves on
any term in $\TRSvartermsp{\TRSone}$:
\begin{eqnarray*}
\TRSonetolambdaII{\varone}&=&\varone;\\
\TRSonetolambdaII{\conone(\termone_1,\ldots,\termone_{\arity{\conone_i}})}&=&\TRSonetolambdaII{\conone_i}
  \TRSonetolambdaII{\termone_1}\ldots\TRSonetolambdaII{\termone_{\arity{\conone_i}}};\\
\TRSonetolambdaII{\funone_i(\termone_1,\ldots,\termone_{\arity{\funone_i}})}&=&\TRSonetolambdaII{\funone_i}
  \TRSonetolambdaII{\termone_1}\ldots\TRSonetolambdaII{\termone_{\arity{\funone_i}}}.
\end{eqnarray*}

\begin{thm}\label{theo:simulcl}
There is a natural number $k$ such that for every
function symbol $\funone$ and for every 
$\termone_1,\ldots,\termone_{\arity{\funone}}\in\TRScontermsp{\TRSone}$,
the following three implications hold, where
$\termtwo$ stands for $\funone(\termone_1,\ldots,\termone_{\arity{\funone}})$
and $\lambdaone$ stands for 
$\TRSonetolambdaII{\funone}\TRSonetolambdaI{\termone_1}\ldots\TRSonetolambdaI{\termone_{\arity{\funone}}}$:
\begin{varitemize}
\item
If $\termtwo$ rewrites to
$\termthree\in\TRScontermsp{\TRSone}$ in $n$ steps, then
$\lambdaone$ rewrites to $\TRSonetolambdaI{\termthree}$ in
at most $kn$ steps.
\item
If $\termtwo$ rewrites to a normal form
$\termthree\notin\TRScontermsp{\TRSone}$, then
$\lambdaone$ rewrites to $\errorterm$.
\item
If $\termtwo$ diverges, then $\lambdaone$ diverges.
\end{varitemize}
\end{thm}
\proof
By an easy combinatorial argument following from the definition
of $\TRSonetolambdaII{\cdot}$. Actually, a slightly stronger statement
should be proved to make the proof formal:
there is a natural number $k$ such that for every $\termtwo\in\TRSvartermsp{\TRSone}$,
for every $\termone_1,\ldots,\termone_m\in\TRScontermsp{\TRSone}$ (where $m$ is 
the number of distinct variables in $\termtwo$),
the following three implications hold, where 
$\lambdaone$ stands for 
$\TRSonetolambdaII{\termtwo}$. 
\begin{varitemize}
\item
If $\termtwo(\termone_1,\ldots,\termone_m)$ rewrites to
$\termthree\in\TRScontermsp{\TRSone}$ in $n$ steps, then
$\lambdaone(\TRSonetolambdaI{\termone_1},\ldots,\TRSonetolambdaI{\termone_m})$ rewrites to 
$\TRSonetolambdaI{\termthree}$ in
at most $kn\length{\termtwo}$ steps.
\item
If $\termtwo(\termone_1,\ldots,\termone_m)$ rewrites to a normal form
$\termthree\notin\TRScontermsp{\TRSone}$, then
$\lambdaone(\TRSonetolambdaI{\termone_1},\ldots,\TRSonetolambdaI{\termone_m})$ 
rewrites to $\errorterm$.
\item
If $\termtwo(\termone_1,\ldots,\termone_m)$ diverges, then 
$\lambdaone(\TRSonetolambdaI{\termone_1},\ldots,\TRSonetolambdaI{\termone_m})$ diverges.
\end{varitemize}
The first statement can be proved by induction on $n$. The second and third one
are quite easy.
\qed
Clearly, the constant $k$ in
Theorem~\ref{theo:simulcl} depends on $\TRSone$, but is independent on the particular
term $\termtwo$.
\subsection{An Example}\label{sect-EsempioTRStoLam}
\newcommand{\TRSadd}{\mathcal{ADD}}
\newcommand{\zero}{\mathbf{0}}
\newcommand{\suc}{\mathbf{s}}
\newcommand{\add}{\mathbf{add}}
In this section, we will describe the encoding of a concrete OCRS called $\TRSadd$ as a set of 
$\lambda$-terms. The signature $\Functions{\TRSadd}$ contains two constructor symbols
$\zero$ and $\suc$, with arity $0$ and $1$ (respectively), and a single function
symbol $\add$ of arity $2$. The only two rules in $\Rules{\TRSadd}$ are the following:
\begin{align*}
\add(\zero,\varone)&\rightarrow\varone;\\
\add(\suc(\varone),\vartwo)&\rightarrow\suc(\add(\varone),\vartwo).
\end{align*}
Let us construct first some $\lambda$-terms in the image of $\TRStolambdaI{\relax}$:
\begin{align*}
\errorterm&=\lambda\varone.\lambda\vartwo.\lambda\varthree.\varthree;\\
\TRStolambdaI{\zero}&=\lambda\varone.\lambda\vartwo.\lambda\varthree.\varone;\\
\TRStolambdaI{\suc(\zero)}&=\lambda\varone.\lambda\vartwo.\lambda\varthree.\vartwo\TRStolambdaI{\zero};\\
\TRStolambdaI{\suc(\suc(\zero))}&=\lambda\varone.\lambda\vartwo.\lambda\varthree.\vartwo\TRStolambdaI{\suc(\zero)};\\
&\vdots\\
\TRStolambdaI{\suc}&=\lambda\varfour.\lambda\varone.\lambda\vartwo.\lambda\varthree.\vartwo\varfour.
\end{align*}
We now take a look at $\TRStolambdaII{\suc}$. By definition:
\begin{align*}
\TRStolambdaII{\suc}&\equiv\conslambda^2_{\varepsilon}\equiv\lambda\vartwo.\vartwo\lambdatwo_{1,2}^0\lambdatwo_{2,2}^0\lambdathree^0_2\\
   &\equiv\lambda\vartwo.\vartwo((\lambda\varone_1.\conslambda^2_{\varone_1})\conslambda^1_{\varepsilon})
                                (\lambda\varthree_1.(\lambda\varone_1.\conslambda^2_{\varone_1})(\conslambda^2_{\varthree_1}))\errorterm.
\end{align*}
This $\lambda$-term indeed ``simulates'' the successor constructor, when fed with an input. Suppose $\termtwo\in\TRScontermsp{\TRSadd}$, then:
\begin{align*}
\TRStolambdaII{\suc}\TRStolambdaI{\zero}&\rightarrow^4(\lambda\varone_1.\conslambda^2_{\varone_1})\conslambda^1_{\varepsilon}
   \equiv(\lambda\varone_1.\conslambda^2_{\varone_1})\TRStolambdaI{\zero}\\
   &\rightarrow\TRStolambdaI{\suc(\zero)};\\
\TRStolambdaII{\suc}\TRStolambdaI{\suc(\termtwo)}&\rightarrow^4(\lambda\varthree_1.(\lambda\varone_1.\conslambda^2_{\varone_1})
  (\conslambda^2_{\varthree_1}))\TRStolambdaI{\termtwo}\rightarrow(\lambda\varthree_1.(\lambda\varone_1.\conslambda^2_{\varone_1})
  \TRStolambdaI{\suc(\termtwo)}\\
  &\rightarrow\TRStolambdaI{\suc(\suc(\termtwo))};\\
\TRStolambdaII{\suc}\errorterm&\rightarrow^4\errorterm.
\end{align*}
Finally, consider $\add$, the only function symbol of $\Functions{\TRSadd}$. By definition:
\begin{align*}
\TRStolambdaII{\add}&\equiv H_1\valueone_1\equiv H_1(\lambda\varone_1.\lambda\vartwo_1.\lambda\vartwo_2.
   \patlambda_{(\zero,\varone),(\suc(\varone),\vartwo)}\vartwo_1\vartwo_2\valuetwo_1^1\valuetwo_i^{2})
\end{align*}
It is easy to verify that, by Lemma~\ref{lemma:pm},
\begin{align*}
\TRStolambdaII{\add}\errorterm\TRStolambdaI{\termone}&\rightarrow^*\errorterm;\\
\TRStolambdaII{\add}\TRStolambdaI{\suc(\zero)}\TRStolambdaI{\suc(\suc(\zero))}&\rightarrow^*\TRStolambdaI{\suc(\suc(\suc(\zero)))}.
\end{align*}
\section{Graph Representation}
\label{Sect:GraphRep}
The previous two sections proved the main simulation result of the paper.
To complete the picture, we show in this section that the unitary cost model
for the (weak call-by-value) $\lambda$-calculus (and hence the number of
rewriting in a OCRSs) is polynomially related to the actual cost of implementing those 
reductions\footnote{As mentioned in the introduction, see~\cite{Sands:Lambda02} for another 
proof of this with other means.}. We do so by introducing term graph rewriting, following~\cite{TGRbarendregt} but
adapting the framework to call-by-value constructor rewriting. Contrarily to what
we did in Section~\ref{sect:CRS}, we will stay abstract here: our attention will
not be restricted to the particular graph rewrite system that is needed to implement 
reduction in the $\lambda$-calculus. 

We refer the reader to our~\cite{DLM09} for more details on efficient simulations between 
term graph rewriting and constructor term rewriting, both under innermost 
(i.e., call-by-value) and outermost (i.e., call-by-name) reduction strategies.

\begin{defi}[Labelled Graph]
Given a signature $\sigone$, a \emph{labelled graph over $\sigone$} consists of a directed
acyclic graph together with an ordering on the outgoing edges of each node and a (partial)
labelling of nodes with symbols from $\sigone$ such that the out-degree of each node
matches the arity of the corresponding symbols (and is $0$ if the labelling is undefined).
Formally, a labelled graph is a triple $\tgone=(\vsone,\ordone,\labelone)$
where: 
\begin{varitemize}
\item
  $\vsone$ is a set of \emph{vertices}.
\item
  $\ordone:\vsone\rightarrow\vsone^*$ is a (total) \emph{ordering function}.
\item
  $\labelone:\vsone\rightharpoonup\vsone$ is a (partial) \emph{labelling function} such
  that the length of $\ordone(\verone)$ is the arity of $\labelone(\verone)$ if
  $\labelone(\verone)$ is defined and is $0$ otherwise.
\end{varitemize}
A labelled graph $(\vsone,\ordone,\labelone)$ is \emph{closed} iff $\labelone$ is a 
total function. 
\end{defi}
Consider the signature $\Sigma=\{\funone,\funtwo,\funthree,\funfour\}$, where
arities of $\funone,\funtwo,\funthree,\funfour$ are $2$, $1$, $0$, $2$ respectively, and
$\funtwo$, $\funthree$, $\funfour$ are constructors. Examples of labelled graphs over 
the signature $\Sigma$ are the following ones:
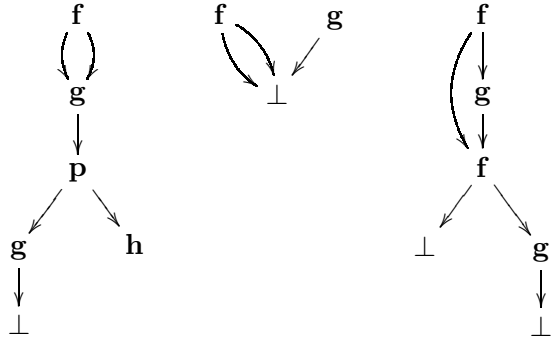
\begin{displaymath}
\xymatrix@R=15pt@C=8pt{
     & \funone \ar@/_/[d]\ar@/^/[d] &   \\
     & \funtwo \ar[d] &   \\
     & \funfour \ar[dl]\ar[dr] &   \\
\funtwo \ar[d] &          & \funthree \\
\bot &          &   \\
}
\hspace{20pt}
\xymatrix@R=15pt@C=8pt{
\funone\ar@/^/[dr]\ar@/_/[dr] & & \funtwo \ar[dl]\\
          & \bot &                         \\
}
\hspace{20pt}
\xymatrix@R=15pt@C=8pt{
     & \funone\ar@/_1pc/[dd]\ar[d] &   \\
     & \funtwo \ar[d] &   \\
     & \funone \ar[dl]\ar[dr] &   \\
\bot &          & \funtwo\ar[d] \\
     &          & \bot  \\
}
\end{displaymath}
The symbol $\bot$ denotes vertices where the underlying labelling function
is undefined (and, as a consequence, no edge departs from such vertices).
Their role is similar to the one of variables in terms.

If one of the vertices of a labelled graph is selected as the \emph{root}, we obtain
a term graph:
\begin{defi}[Term Graph]
A \emph{term graph}, is a quadruple 
$\tgone=(\vsone,\ordone,\labelone,\rootone)$, where 
$(\vsone,\ordone,\labelone)$ is a labelled graph and
$\rootone\in\vsone$ is the \emph{root} of the term graph.
\end{defi}
The following are graphic representations of some term graphs.
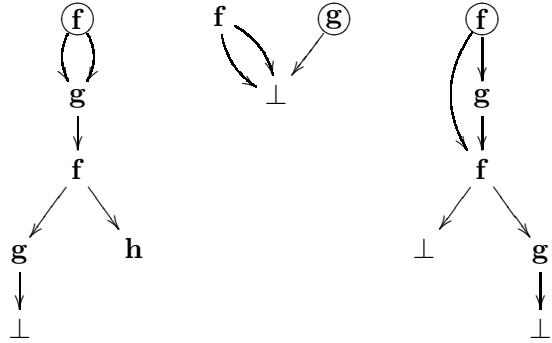
\begin{displaymath}
\xymatrix@R=15pt@C=8pt{
     &  *+[o][F]{\funone} \ar@/_/[d]\ar@/^/[d] &   \\
     & \funtwo \ar[d] &   \\
     & \funone \ar[dl]\ar[dr] &   \\
\funtwo \ar[d] &          & \funthree \\
\bot &          &   \\
}
\hspace{20pt}
\xymatrix@R=15pt@C=8pt{
 \funone\ar@/^/[dr]\ar@/_/[dr] & &  *+[o][F]{\funtwo} \ar[dl]\\
          & \bot &                         \\
}
\hspace{20pt}
\xymatrix@R=15pt@C=8pt{
     & *+[o][F]{\funone}\ar@/_1pc/[dd]\ar[d] &   \\
     & \funtwo \ar[d] &   \\
     & \funone \ar[dl]\ar[dr] &   \\
\bot &          & \funtwo\ar[d] \\
     &          & \bot  \\
}
\end{displaymath}
The root is the only vertex drawn inside a circle.

There are some classes of paths which are particularly relevant for our purposes.
\begin{defi}[Path]
A path $\verone_1,\ldots,\verone_n$ in a labelled graph
$\tgone=(\vsone,\ordone,\labelone)$ is said to be:
\begin{varitemize}
\item
  A \emph{constructor path} iff for every $1\leq i\leq n$, the symbol
  $\labelone(\verone_i)$ is a constructor;
\item
  A \emph{pattern path} iff for every $1\leq i\leq n$, 
  $\labelone(\verone_i)$ is either a constructor symbol or is undefined;
\item
  A \emph{left path} iff $n\geq 1$, the symbol $\labelone(\verone_1)$ is 
  a function symbol and $\verone_2,\ldots,\verone_n$
  is a pattern path.
\end{varitemize}
\end{defi}

\begin{defi}[Homomorphism]
A \emph{homomorphism} between two labelled graphs 
$\tgone=(\vsone_\tgone,\ordone_\tgone,\labelone_\tgone)$ and 
$\tgtwo=(\vsone_\tgtwo,\ordone_\tgtwo,\labelone_\tgtwo)$ over the same 
signature $\sigone$ is a function $\homone$ from $\vsone_\tgone$ to $\vsone_\tgtwo$ 
preserving the term graph structure. In particular
\begin{eqnarray*}
  \labelone_\tgtwo(\homone(\verone))&=&\labelone_\tgone(\verone)\\
  \ordone_\tgtwo(\homone(\verone))&=&\homone^*(\ordone_\tgone(\verone))
\end{eqnarray*}
for any $\verone\in\domain{\labelone}$, where $\homone^*$ is the obvious
generalization of $\homone$ to sequences of vertices. 
An \emph{homomorphism} between two term graphs 
$\tgone=(\vsone_\tgone,\ordone_\tgone,\labelone_\tgone,\rootone_\tgone)$ and 
$\tgtwo=(\vsone_\tgtwo,\ordone_\tgtwo,\labelone_\tgtwo,\rootone_\tgtwo)$ is
a homomorphism between $(\vsone_\tgone,\ordone_\tgone,\labelone_\tgone)$ and 
$(\vsone_\tgtwo,\ordone_\tgtwo,\labelone_\tgtwo)$ such that
$\homone(\rootone_\tgone)=\rootone_\tgtwo$. Two labelled graphs $\tgone$ and $\tgtwo$ are 
isomorphic iff there is a bijective homomorphism from
$\tgone$ to $\tgtwo$; in this case, we write $\tgone\cong\tgtwo$. Similarly
for term graphs.
\end{defi}
In the following, we will consider term graphs modulo isomorphism, i.e., $\tgone=\tgtwo$
iff $\tgone\cong\tgtwo$. Observe that two isomorphic term graphs have the same graphical
representation.
\begin{defi}[Graph Rewrite Rule]
A \emph{graph rewrite rule} over a signature $\sigone$ 
is a triple $\rrone=(\tgone,\rootone,\roottwo)$ such that:
\begin{varitemize}
\item
  $\tgone$ is a labelled graph;
\item
  $\rootone,\roottwo$ are vertices of $\tgone$, called 
  the \emph{left root} and the \emph{right root} of $\rrone$,
  respectively.
\item
  Any path starting in $\rootone$ is a
  left path.
\end{varitemize}
\end{defi}
The following are examples of graph rewrite rules, assuming $a$ to be a function
symbol and $b,c,d$ to be constructors:
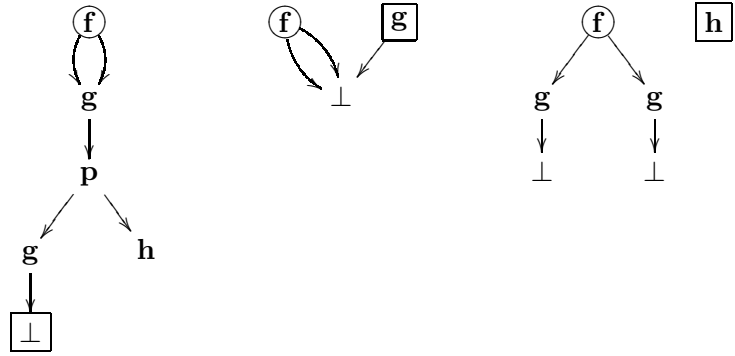
\begin{displaymath}
\xymatrix@R=15pt@C=8pt{
     &  *+[o][F]{\funone} \ar@/_/[d]\ar@/^/[d] &   \\
     & \funtwo \ar[d] &   \\
     & \funfour \ar[dl]\ar[dr] &   \\
\funtwo \ar[d] &          & \funthree \\
*+[F]{\bot} &          &   \\
}
\hspace{40pt}
\xymatrix@R=15pt@C=8pt{
 *+[o][F]{\funone}\ar@/^/[dr]\ar@/_/[dr] & &  *+[F]{\funtwo} \ar[dl]\\
          & \bot &                         \\
}
\hspace{40pt}
\xymatrix@R=15pt@C=8pt{
        & *+[o][F]{\funone}\ar[dl]\ar[dr] &  &  *+[F]{\funthree}\\
\funtwo\ar[d] &                  & \funtwo\ar[d] &   \\
\bot    &                  & \bot    &   \\
}
\end{displaymath}
\begin{defi}[Subgraph]
Given a labelled graph $\tgone=(\vsone_\tgone,\ordone_\tgone,\labelone_\tgone)$ 
and any vertex $\verone\in\vsone_\tgone$, the \emph{subgraph of $\tgone$ rooted
at $\verone$}, denoted $\subgr{\tgone}{\verone}$, is the
term graph $(\vsone_{\subgr{\tgone}{\verone}},\ordone_{\subgr{\tgone}{\verone}},\labelone_{\subgr{\tgone}{\verone}},
\rootone_{\subgr{\tgone}{\verone}})$ where
\begin{varitemize}
\item
  $\vsone_{\subgr{\tgone}{\verone}}$ is the subset of $\vsone_\tgone$ whose elements
  are vertices which are reachable from $\verone$ in $\tgone$.
\item
  $\ordone_{\subgr{\tgone}{\verone}}$ and $\labelone_{\subgr{\tgone}{\verone}}$ are
  the appropriate restrictions of $\ordone_\tgone$ and $\labelone_\tgone$
  to $\vsone_{\subgr{\tgone}{\verone}}$.
\item
  $\rootone_{\subgr{\tgone}{\verone}}$ is $\verone$.
\end{varitemize}
\end{defi}
\begin{defi}[Redex]
Given a labelled graph $\tgone$, a \emph{redex} for $\tgone$ is
a pair $(\rrone,\homone)$, where $\rrone$ is a rewrite rule 
$(\tgtwo,\rootone,\roottwo)$ and $\homone$ is a homomorphism
between $\subgr{\tgtwo}{\rootone}$ and $\tgone$
such that for any vertex $\verone\in\vsone_{\subgr{\tgtwo}{\rootone}}$
with $\verone\notin\domain{\labelone_{\subgr{\tgtwo}{\rootone}}}$, 
any path starting in $\homone(\verone)$ is a constructor path.
\end{defi}
The last condition in the definition of a redex is needed to
capture the call-by-value nature of the rewriting process.

Given a term graph $\tgone$ and a redex $((\tgtwo,\rootone,\roottwo),\homone)$,
the result of firing the redex is another term graph obtained by
successively applying the following three steps to $\tgone$:
\begin{enumerate}[1.]
\item
  The \emph{build phase}: create an isomorphic copy of the portion of
  $\subgr{\tgtwo}{\roottwo}$ not contained in
  $\subgr{\tgtwo}{\rootone}$, and add it to $\tgone$, obtaining $\tgthree$.
  The underlying ordering and labelling functions are defined in the natural
  way.
\item
  The \emph{redirection phase}: all edges in $\tgthree$ pointing to $\homone(\rootone)$
  are replaced by edges pointing to the copy of $\roottwo$. If $\homone(\rootone)$
  is the root of $\tgone$, then the root of the newly created graph will be
  the newly created copy of $\roottwo$. The graph $\tgfour$ is obtained.
\item
  The \emph{garbage collection phase}: all vertices which are not accessible
  from the root of $\tgfour$ are removed. The graph $\tgfive$ is obtained.
\end{enumerate}
We will write $\tgone\rewrite{(\tgtwo,\rootone,\roottwo)}\tgfive$ (or simply
$\tgone\rewrgraph\tgfive$, if this does not cause ambiguity) in this case.

As an example, consider the term graph $\tgone$ and the rewrite rule 
$\rrone=(\tgtwo,\rootone,\roottwo)$:
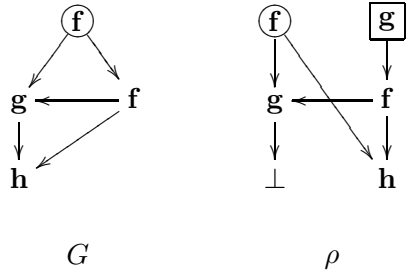
\begin{displaymath}
\xymatrix@R=15pt@C=8pt{
  &  *+[o][F]{\funone}\ar[dl]\ar[dr] &   \\
  \funtwo \ar[d] & & \funone \ar[ll]\ar[dll] \\
  \funthree & &  \\
  & \tgone & \\
}
\hspace{40pt}
\xymatrix@R=15pt@C=8pt{
 *+[o][F]{\funone} \ar[d]\ar[ddrr] & & *+[F]{\funtwo}\ar[d] \\
 \funtwo \ar[d] & & \funone\ar[ll]\ar[d] \\
\bot & & \funthree \\
 & \rrone & \\
}
\end{displaymath}
There is a homomorphism $\homone$ from
$\subgr{\tgtwo}{\rootone}$ to
$\tgone$. In particular, $\homone$ maps $\rootone$ to
the rightmost vertex in $\tgone$.
Applying the build phase and the redirection phase we get $\tgthree$
and $\tgfour$ as follows:
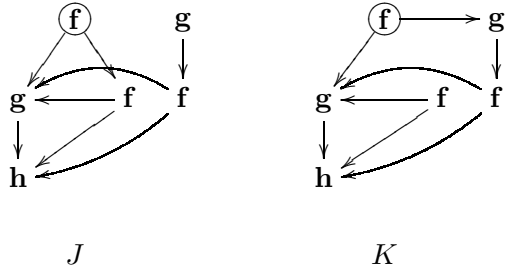
\begin{displaymath}
\xymatrix@R=15pt@C=8pt{
  &  *+[o][F]{\funone}\ar[dl]\ar[dr] & & \funtwo\ar[d]  \\
  \funtwo \ar[d] & & \funone \ar[ll]\ar[dll] & \funone\ar@/^/[dlll]\ar@/_1pc/[lll] \\
  \funthree & & & \\
  & \tgthree & & \\
}
\hspace{40pt}
\xymatrix@R=15pt@C=8pt{
  &  *+[o][F]{\funone}\ar[dl]\ar[rr] & & \funtwo\ar[d]  \\
  \funtwo \ar[d] & & \funone\ar[ll]\ar[dll] & \funone\ar@/^/[dlll]\ar@/_1pc/[lll] \\
  \funthree & & & \\
  & \tgfour & & \\
}
\end{displaymath}
Finally, applying the garbage collection phase, we get the
result of firing the redex $(\rrone,\homone)$:
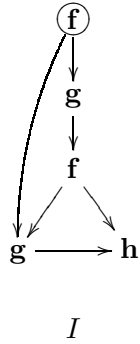
\begin{displaymath}
\xymatrix@R=15pt@C=8pt{
 & *+[o][F]{\funone}\ar@/_/[dddl]\ar[d] & \\
 & \funtwo\ar[d] & \\
 & \funone\ar[dl]\ar[dr] & \\
 \funtwo\ar[rr] & & \funthree\\
 & I & \\
}
\end{displaymath}

\begin{defi}
A constructor graph rewrite system (CGRS) over a signature $\sigone$ consists of
a set $\sgrone$ of graph rewrite rules  on $\sigone$.
\end{defi}
\subsection{From Term Rewriting to Graph Rewriting}\label{Sect-TRtoGR}
Any term $\termone$ over a signature $\sigone$ can be turned into a
graph $\CtoCG{\termone}$ in the obvious way: take as $\CtoCG{\termone}$ the abstract
syntax tree of $\termone$, where vertices are in one-to-one correspondence
with symbol occurrences in $\termone$. Conversely, any term graph
$\tgone$ over $\sigone$ can be turned into a term $\CGtoC{\tgone}$
over $\sigone$ by simply unfolding the graph, that is applying (the label of) any vertex
to (the terms obtained as unfolding of) its sons (remember: we only consider acyclic graphs here).
We omit the boring formal definitions of both $\CtoCG{\cdot}$ and $\CGtoC{\cdot}$; it is clear that 
for any term $\termone$, $\CGtoC{\CtoCG{\termone}}=\termone$, while in general 
$\CtoCG{\CGtoC{\tgone}}$ is not equal to $\tgone$, since the sharing present in $\tgone$ is lost during the unfolding. 

\begin{defi}
Given a constructor rewriting system $\srrone$ over $\sigone$, the 
corresponding constructor graph rewriting system $\CtoCG{\srrone}$ is defined 
by translating the terms with $\CtoCG{\cdot}$ and by translating any term 
rewrite rule $\termone\rewrTRS\termtwo$ 
over $\sigone$ into a graph rewrite rule $(\tgone,\rootone,\roottwo)$
as follows:
\begin{varitemize}
\item
  Take the graphs $\CtoCG{\termone}$ and $\CtoCG{\termtwo}$ (which are
  trees, in fact). 
\item
  From the union of these two trees, share those
  nodes representing the same variable in $\termone$ and $\termtwo$.
  This is $\tgone$.
\item
  Take $\rootone$ to be the root of $\termone$ in $\tgone$ and
  $\roottwo$ to be the root of $\termtwo$ in $\tgone$.
\end{varitemize}
\end{defi}

As an example, consider the rewrite rule
$$
\funone(\funtwo(\varone),\vartwo)\rewrTRS \funtwo(\funone(\vartwo,\funone(\vartwo,\varone))).
$$
Its translation as a graph rewrite rule is the following:
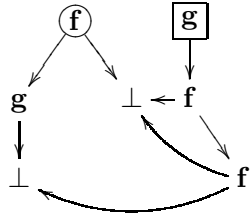
\begin{displaymath}
\xymatrix@R=15pt@C=8pt{
 & *+[o][F]{\funone}\ar[dl]\ar[dr] & & *+[F]{\funtwo}\ar[d] & \\
 \funtwo\ar[d] & & \bot & \funone\ar[l]\ar[dr] & \\
 \bot & & & & \funone\ar@/^/[ull]\ar@/^1pc/[llll]
}\vspace{12 pt}
\end{displaymath}

Given a constructor rewriting system $\srrone$, it is easy to realize that
the following invariant is preserved while performing
rewriting in $\CtoCG{\srrone}$: whenever any vertex $\verone$ can
be reached by two distinct paths starting at the root
(i.e., $\verone$ is \emph{shared}), any path starting at
$\verone$ is a constructor path. A term graph
satisfying this invariant is said to be 
\emph{constructor-shared}.

Constructor-sharedness holds for term graphs coming from terms and
is preserved by graph rewriting:
\begin{lem}\label{lemma:constructorsharedness}
For every closed term $\termone$, $\CtoCG{\termone}$ is constructor-shared.
Moreover, if $\tgone$ is closed and constructor-shared and $\tgone\rewrgraph\tgfive$ in $\CtoCG{\srrone}$, 
then $\tgfive$ is constructor-shared.
\end{lem}
\proof
The fact that $\CtoCG{\termone}$ is constructor-shared for every $\termone$ follows
from the way the $\CtoCG{\cdot}$ map is defined: it does not introduce any
sharing. Now, suppose $\tgone$ is constructor-shared and
$$
\tgone\rewrite{(\tgtwo,\rootone,\roottwo)}\tgfive
$$
where $(\tgtwo,\rootone,\roottwo)$ corresponds to a term rewrite
rule $\termone\rewrTRS\termtwo$. The term graph $\tgthree$ obtained from
$\tgone$ by the build phase is itself constructor-shared: it is obtained from
$\tgone$ by adding some new nodes, namely an isomorphic copy of the portion
of $\subgr{\tgtwo}{\roottwo}$ not contained in $\subgr{\tgtwo}{\rootone}$. Notice
that $\tgthree$ is constructor-shared in a stronger sense: any vertex which
can be reached from the newly created copy of $\roottwo$ by two distinct paths
must be a constructor path. This is a consequence of $(\tgtwo,\rootone,\roottwo)$ 
being a graph rewrite rule corresponding to a term rewrite
rule $\termone\rewrTRS\termtwo$, where the only shared vertices are those
where the labelling function is undefined. The redirection phase preserves
itself constructor-sharedness, because only one pointer is redirected
(the vertex is labelled by a function symbol) and the destination 
of this redirection is a vertex (the newly created
copy of $\roottwo$) which had no edge incident to it. Clearly, the
garbage collection phase preserves constructor-sharedness.
\qed
\begin{lem}\label{lemma:NFconstructorshared}
A closed term graph $\tgone$ in  $\CtoCG{\srrone}$ is a 
normal form iff $\CGtoC{\tgone}$ is a normal form.
\end{lem}
\proof
Clearly, if a closed term graph $\tgone$ is in normal form,
then $\CGtoC{\tgone}$ is a term in normal form, because each
redex in $\tgone$ translates to a redex in $\CGtoC{\tgone}$.
On the other hand, if $\CGtoC{\tgone}$ is in normal form,
then $\tgone$ is in normal form: each redex in $\CGtoC{\tgone}$
translates back to a redex in $\tgone$.
\qed
Reduction at the level of graphs correctly simulates reduction
at the level of terms, but only if the underlying graphs
are constructor shared:
\begin{lem}\label{lemma:CGtoC}
If $\tgone$ is closed and constructor-shared, and 
$\tgone\rewrgraph\tgfive$ in  $\CtoCG{\srrone}$, then
$\CGtoC{\tgone}\rewrgraph\CGtoC{\tgfive}$ in $\srrone$.
\end{lem}
\proof
The fact that each reduction step starting in $\tgone$ can be
mimicked in $\CGtoC{\tgone}$ is known
from the literature. If $\tgone$ is constructor-shared,
then the simulation is done in exactly one reduction step,
because any redex in a constructor-shared
term graph cannot be shared.
\qed
When $\tgone$ in not constructor-shared, a counterexample
can be easily built. Consider the term rewrite rule
$\funone(\funthree,\funthree)\rewrTRS\funthree$ and the following term graph:
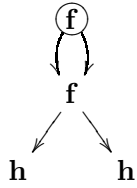
\begin{displaymath}
\xymatrix@R=15pt@C=8pt{
 & *+[o][F]{\funone}\ar@/_/[d]\ar@/^/[d] &  \\
 & \funone\ar[dl]\ar[dr] & \\
 \funthree & & \funthree\\
}
\end{displaymath}
It corresponds to $\funone(\funone(\funthree,\funthree),\funone(\funthree,\funthree))$, and it is not
constructor-shared, since the shared vertex $a$ is not a constructor. 
It rewrites in \emph{one} step to 
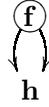
\begin{displaymath}
\xymatrix@R=15pt@C=8pt{
 & *+[o][F]{\funone}\ar@/_/[d]\ar@/^/[d] &  \\
 & \funthree & \\
}
\end{displaymath}
while the term $\funone(\funone(\funthree,\funthree),\funone(\funthree,\funthree))$ rewrites to $\funone(\funthree,\funthree)$ in
\emph{two} steps.

As can be expected, graph reduction is also complete with respect to
term reduction, with the only \emph{proviso} that term graphs must
be constructor-shared:
\begin{lem}\label{lemma:CtoCG}
If $\termone\rewrTRS\termtwo$ in $\srrone$, $\tgone$ is constructor-shared 
and $\CGtoC{\tgone}=\termone$, then
$\tgone\rewrgraph\tgfive$ in $\CtoCG{\srrone}$, where $\CGtoC{\tgfive}=\termtwo$. \qed
\end{lem}

\begin{thm}[Graph Reducibility]\label{theo:graphreducible}
For every constructor rewrite system $\srrone$ over $\sigone$ and
for every term $\termone$ over $\sigone$, the following two conditions are
equivalent:
\begin{enumerate}[\em 1.]
\item
  $\termone\rewrTRS^n\termtwo$ in $\srrone$, where $\termtwo$ is in normal form;
\item
  $\CtoCG{\termone}\rewrTRS^n\tgone$ in $\CtoCG{\srrone}$, where $\tgone$ is in normal
  form and $\CGtoC{\tgone}=\termtwo$.
\end{enumerate}
\end{thm}
\proof
Suppose $\termone\rewrgraph^n\termtwo$, where $\termtwo$ is in normal form. 
Then, by applying Lemma~\ref{lemma:CtoCG}, we obtain a term graph $\tgone$ such that
$\CtoCG{\termone}\rewrgraph^n\tgone$ and $\CGtoC{\tgone}=\termtwo$.
By Lemma~\ref{lemma:constructorsharedness}, $\tgone$ is constructor-shared and, by 
Lemma~\ref{lemma:NFconstructorshared}, 
it is in normal form. Now, suppose $\CtoCG{\termone}\rewrgraph^n\tgone$ where
$\CGtoC{\tgone}=\termtwo$ and $\tgone$ is in normal form. By
applying $n$ times Lemma~\ref{lemma:CGtoC}, we obtain
that $\CGtoC{\CtoCG{\termone}}\rewrTRS^n\CGtoC{\tgone}=\termtwo$.
But $\CGtoC{\CtoCG{\termone}}=\termone$ and $\termtwo$ is a normal form 
by Lemma~\ref{lemma:NFconstructorshared}, since 
$\CtoCG{\termone}$ and $\tgone$ are
constructor shared due to Lemma~\ref{lemma:constructorsharedness}.
\qed

There are \emph{term} rewrite systems which are not graph reducible, i.e.
for which the two conditions of Theorem~\ref{theo:graphreducible} are
not equivalent (see~\cite{TGRbarendregt}). However, any 
\emph{orthogonal constructor} rewrite system is graph reducible, due to the 
strict constraints on the shape of rewrite rules~\cite{Plump90ggacs}.
This result can be considered as a by-product of our analysis, for which graph rewriting
is only instrumental.

\subsection{Lambda-Terms Can Be Efficiently Reduced by Graph Rewriting}
\label{sect:MainResult}
As a corollary of Theorems~\ref{theo:graphreducible} and~\ref{theo:termreducible},
we may reduce $\lambda$-terms using term graphs.
To this purpose, we apply the construction of the previous section
to the OCRS $\TRS$ that we defined in Section~\ref{sect:CRS}. Let 
then $\GRS=\CtoCG{\TRS}$:
\begin{cor}
Let $\lambdaone\in\Lambdaterms$ be a closed $\lambda$-term. The following
two conditions are equivalent:
\begin{enumerate}[\em 1.]
\item
  $\lambdaone\rewrlambdav^n\lambdatwo$ where $\lambdatwo$ is in normal form;
\item
  $\TRStoGRS{\LambdatoTRS{\lambdaone}}\rewrTRS^n\tgone$ where
  $\TRSonetolambdaI{\GRStoTRS{\tgone}}=\lambdatwo$ and $\tgone$ is in normal form. \qed
\end{enumerate}
\end{cor} 
Let us now analyze more closely the combinatorics 
of graph rewriting in $\GRS$, so that we can obtain information on the efficiency of this simulation. 
\begin{varitemize}
\item
  Consider a closed $\lambda$-term $\lambdaone$ and a term graph $\tgone$ such
  that $\TRStoGRS{\LambdatoTRS{\lambdaone}}\rewrgraph^*\tgone$. By Proposition~\ref{prop:constred}
  and Lemma~\ref{lemma:CGtoC}, for every constructor $\constr{\varone}{\lambdatwo}$ appearing
  as a label of a vertex in $\tgone$, $\lambdatwo$ is a subterm of $\lambdaone$. 
\item
  As a consequence, if $\TRStoGRS{\LambdatoTRS{\lambdaone}}\rewrgraph^*\tgone\rewrgraph\tgtwo$,
  then the difference $\length{\tgtwo}-\length{\tgone}$ cannot be too big: at most
  $\length{\lambdaone}$. Therefore, if 
  $\TRStoGRS{\LambdatoTRS{\lambdaone}}\rewrgraph^n\tgone$ then $\length{\tgone}\leq(n+1)\length{\lambdaone}$.
  Here, we exploit in an essential way the possibility of sharing constructors.
\item
  Whenever $\TRStoGRS{\LambdatoTRS{\lambdaone}}\rewrgraph^n\tgone$, computing
  a graph $\tgtwo$ such that $\tgone\rewrgraph\tgtwo$ takes polynomial time in
  $\length{\tgone}$, which is itself polynomially bounded by $n$ and $\length{\lambdaone}$.
\end{varitemize}
Hence (recall that $\Time{\lambdaone}$ is the number of weak call-by-value beta steps to normal form): 
\begin{thm}\label{thm:main}
There is a polynomial $p:\N^2\rightarrow\N$ such that for every $\lambda$-term $\lambdaone$,
the normal form of $\TRStoGRS{\LambdatoTRS{\lambdaone}}$ can be computed in time at most 
$p(|\lambdaone|,\Time{\lambdaone})$. \qed
\end{thm}
This cannot be achieved when using explicit representations
of $\lambda$-terms. Moreover, reading back a $\lambda$-term from a term graph can take exponential
time.

We can complement Theorem~\ref{thm:main} with a completeness statement --- any universal computational model 
with an invariant cost model can be embedded in the $\lambda$-calculus with a polynomial
overhead. We can exploit for this the analogous result we proved in~\cite{CIE2006} (Section 4, Theorem 1) --- 
the unitary cost model is easily proved to be more parsimonious than 
the difference cost model considered in~\cite{CIE2006}.

\begin{thm}
Let $f:\Sigma^*\rightarrow\Sigma^*$ be computed by a Turing machine
$\TMone$ in time $g$. Then, there are a $\lambda$-term $\lambdatwo_\TMone$
and a suitable encoding $\cod{\cdot}:\Sigma^*\rightarrow\Lambdaterms$ 
such that $\lambdatwo_\TMone\cod{v}$ normalizes to $\cod{f(v)}$ in 
$O(g(|v|))$ beta steps. \qed
\end{thm}

The encoding $\cod{\cdot}$ mentioned in the theorem depends only on (the cardinality of) 
$\Sigma$ (but not on the Turing machine). 
Interestingly enough it exploits once again the scheme that we used in Definition~\ref{Def-TRSonetolambdaI}: encode the empty string $\varepsilon$ as a zero-ary constructor,
and any symbol in $\Sigma$ as a unary constructor (see~\cite{CIE2006} for details).
\section{Variations: Call-by-Name Reduction}
\label{sect:HeadReduction}
In the previous sections, $\lambda$-calculus was endowed with weak call-by-value reduction.
The same technique, however, can be applied to weak call-by-name reduction, as we will sketch in
this section. $\Lambdaterms$ is now endowed with a relation $\rewrlambdah$ defined
as follows:
$$
\begin{array}{ccccc}
  \infer{(\lambda\varone.\lambdaone)\lambdatwo\rewrlambdah\lambdaone\{\lambdatwo/\varone\}}{}
  & & & &
  \infer{\lambdaone\lambdathree\rewrlambdah\lambdatwo\lambdathree}{\lambdaone\rewrlambdah\lambdatwo}
\end{array}
$$
Similarly to the call-by-value case, $\Timew{\lambdaone}$ stands for the number of
reduction steps to the normal form of $\lambdaone$ (if any). Since the relation
$\rewrlambdah$ is deterministic (i.e., functional), $\Timew{\lambdaone}$ is well-defined.

We need another OCRS, called $\TRSW$, which is similar to $\TRS$ but designed
to simulate weak call-by-name reduction:
\begin{varitemize}
\item
  The signature $\Functions{\TRSW}$ includes the binary function symbol $\appTRS$ and
  constructor symbols $\constr{\varone}{M}$ for every $\lambdaone\in\Lambdaterms$ and
  every $\varone\in\Variables$, exactly as $\Functions{\TRS}$. Moreover, there is another
  binary constructor symbol $\cappTRSW$. 
  To every term $\lambdaone\in\Lambdaterms$ we can associate
  terms $\LambdatoTRSWaux{\lambdaone},\LambdatoTRSW{\lambdaone}\in\TRSWvarterms$ as follows:
  \begin{eqnarray*}
    \LambdatoTRSWaux{\varone}&=&\varone\\
    \LambdatoTRSWaux{\lambda\varone.\lambdaone}&=&\constr{\varone}{\lambdaone}(\varone_1,\ldots,\varone_{n}),
      \mbox{ where $\FV{\lambda\varone.\lambdaone}=\varone_1,\ldots,\varone_n$}\\
    \LambdatoTRSWaux{\lambdaone\lambdatwo}&=&\cappTRSW(\LambdatoTRSWaux{\lambdaone},\LambdatoTRSWaux{\lambdatwo})\\
    \LambdatoTRSW{\varone}&=&\varone\\
    \LambdatoTRSW{\lambda\varone.\lambdaone}&=&\constr{\varone}{\lambdaone}(\varone_1,\ldots,\varone_{n}),
      \mbox{ where $\FV{\lambda\varone.\lambdaone}=\varone_1,\ldots,\varone_n$}\\
    \LambdatoTRSW{\lambdaone\lambdatwo}&=&\appTRS(\LambdatoTRSW{\lambdaone},\LambdatoTRSWaux{\lambdatwo})
  \end{eqnarray*}
  Notice that $\LambdatoTRSWaux{\cdot}$ maps $\lambda$-terms to \emph{constructor} terms, while
  terms obtained via $\LambdatoTRSW{\cdot}$ can contain function symbols. The ``official'' translation of a term $M$ is thus $\LambdatoTRSW{M}$, where only the applications ``on the spine'' of $M$ are  encoded with $\appTRS$. All other applications are frozen by the constructor $\cappTRSW$.
\item
    The rewrite rules in $\Rules{\TRSW}$ are all the rules in the following form:
    \begin{eqnarray*}
      \appTRS(\constr{\varthree}{\varthree},\cappTRSW(\varfour,\varfive))&\rewrTRSW&\appTRS(\varfour,\varfive)\\
      \appTRS(\constr{\varthree}{\varthree},\constr{\varone}{\lambdaone}(\varone_1,\ldots,\varone_n))&\rewrTRSW&
        \constr{\varone}{\lambdaone}(\varone_1,\ldots,\varone_n)\\
      \appTRS(\constr{\varthree}{\varfour}(\cappTRSW(\varfive,\varsix)),\varseven)&\rewrTRSW&\appTRS(\varfive,\varsix)\\
      \appTRS(\constr{\varthree}{\varfour}(\constr{\varone}{\lambdaone}(\varone_1,\ldots,\varone_n)),\varseven)&\rewrTRSW&
        \constr{\varone}{\lambdaone}(\varone_1,\ldots,\varone_n)\\
     \appTRS(\constr{\vartwo}{\lambdatwo}(\vartwo_1,\ldots,\vartwo_{m}),\vartwo)&\rewrTRSW&\LambdatoTRSW{\lambdatwo}
    \end{eqnarray*}
    where $\lambdaone$ ranges over $\lambda$-terms, $\lambdatwo$ ranges over abstractions and
    applications,
    $\FV{\lambda\varone.\lambdaone}=\varone_1,\ldots,\varone_n$
    and $\FV{\lambda\vartwo.\lambdatwo}=\vartwo_1,\ldots,\vartwo_m$. These rewrite rules
    are said to be \emph{ordinary rules}. We also need the following \emph{administrative} rule:
    $$
    \appTRS(\cappTRSW(\varone,\vartwo),\varthree)\rewrTRSW\appTRS(\appTRS(\varone,\vartwo),\varthree).
    $$
\end{varitemize}
The CTRS $\TRSW$ is more complicated than $\TRS$, because we need to force reduction to happen only in head position. The applications $\appTRS$ (on the spine) may be fired immediately.
Observe, however, that the main rewriting rule (the last of the ordinary ones) is restricted to those 
$\constr{\vartwo}{\lambdatwo}$
where $\lambdatwo$ is \emph{not} a variable. When $\lambdatwo$ is a single variable, 
the corresponding beta redex would be either $(\lambda x.x)\lambdathree$ or 
$(\lambda z.w)\lambdathree$, with $w$ free. In the former case, an application at the top
level of $\lambdathree$ (encoded as a $\cappTRSW$ at this point) would become the top level
application of the spine of the reduct: the first ordinary reduction rule handles this case, unfreezing
$\cappTRSW$ into  $\appTRS$. When, on the other hand, the encoded redex is 
$(\lambda z.w)\lambdathree$, we do not need to worry for $\lambdathree$, which will be discarded,
but in the term-reduction we must take care of the eventual substitution that may occur for
$w$: the term substituted for $w$ may have a top level frozen application $\cappTRSW$ that must be
converted into an $\appTRS$ --- this is the role of the third ordinary reduction rule. 
The second and fourth reduction rules just handle the remaining cases (they would be
instances of the last ordinary rule if this was not restricted to the non-variable cases).
A last remark on the administrative rule. There are never administrative redexes in the translation $\LambdatoTRSW{\lambdaone}$ of a term. During reduction, however, by the effect of the other rules a frozen application (a $\cappTRSW$)
    may appear on the spine. The administrative rule recognizes this situation and unfreezes the application.

As usual, to every term $\termone\in\TRSWvarterms$ we can associate a term $\TRSWtolambda{\termone}$:
\begin{eqnarray*}
  \TRSWtolambda{\varone}&=&\varone\\
  \TRSWtolambda{\appTRS(\termtwo,\termthree)}=\TRSWtolambda{\cappTRSW(\termtwo,\termthree)}
     &=&\TRSWtolambda{\termtwo}\TRSWtolambda{\termthree}\\
  \TRSWtolambda{\constr{\varone}{\lambdaone}(\termone_1,\ldots\termone_n)}&=&
  (\lambda\varone.\lambdaone)\{\TRSWtolambda{\termone_1}/\varone_1,\ldots,\TRSWtolambda{\termone_n}/\varone_n\}
\end{eqnarray*}
where $\FV{\lambda\varone.\lambdaone}=\varone_1,\ldots,\varone_n$.
A term $\termone\in\TRSWterms$ is \emph{canonical} if either $\termone=\constr{\varone}{\lambdaone}(\termone_1\,\ldots,\termone_n)\in\TRSWconterms$ or
$\termone=\appTRS(\termtwo,\termthree)$ where $\termtwo$ is canonical and $\termthree\in\TRSWconterms$.
\begin{lem}\label{lemma:closedcanonical}
For every closed $\lambdaone\in\Lambdaterms$, $\LambdatoTRSW{\lambdaone}$ is canonical.
\end{lem}
\proof
By a straightforward induction on $\lambdaone$.
\qed
The obvious variation on Equation~(\ref{equat:commute}) holds here:
\begin{equation}\label{equat:commutew}
\TRSWtolambda{\LambdatoTRSW{\lambdaone}\{\termone_1/\varone_1,\ldots,\termone_n/\varone_n\}}=
\lambdaone\{\TRSWtolambda{\termone_1}/\varone_1,\ldots,\TRSWtolambda{\termone_n}/\varone_n\}.
\end{equation}
$\TRSW$ mimics call-by-name reduction in much the same way $\TRS$ mimics call-by-value
reduction. However, one reduction step in the $\lambda$-calculus corresponds to $n\geq 1$ steps
in $\TRSW$, although $n$ is kept under control:
\begin{lem}\label{lemma:simul}
Suppose that $\termone\in\TRSWterms$ is canonical and that $\termone\rewrTRSW\termtwo$. Then
there is a natural number $n$ such that:
\begin{enumerate}[\em 1.]
\item
  $\TRSWtolambda{\termone}\rewrlambdah\TRSWtolambda{\termtwo}$;
\item
  There is a canonical term $\termthree\in\TRSWterms$ such that $\termtwo\rewrTRSW^n\termthree$;
\item
  $\plength{\termfour}{\appTRS}=\plength{\termtwo}{\appTRS}+m$ whenever $\termtwo\rewrTRSW^m\termfour$
  and $m\leq n$;
\item
  $\TRSWtolambda{\termfour}=\TRSWtolambda{\termtwo}$ whenever $\termtwo\rewrTRSW^m\termfour$
  and $m\leq n$.
\end{enumerate}
\end{lem}
\proof
A term $\termone$ is said to be \emph{semi-canonical} iff $\termone=\appTRS(\termtwo,\termthree)$, where
$\termthree\in\TRSWconterms$ and $\termtwo$ is either semi-canonical or is itself an element
of $\TRSWconterms$. We now prove that if $\termone$ is semi-canonical, there there are a natural
number $n$ and a canonical term $\termtwo$ such that:
\begin{varitemize}
\item
  $\termone\rewrTRSW^n\termtwo$;
\item
  $\plength{\termthree}{\appTRS}=\plength{\termone}{\appTRS}+m$ whenever $\termone\rewrTRSW^m\termthree$
  and $m\leq n$;
\item
  $\TRSWtolambda{\termthree}=\TRSWtolambda{\termone}$ whenever $\termone\rewrTRSW^m\termthree$
  and $m\leq n$.
\end{varitemize}
We proceed by induction on $\length{\termone}$. By definition $\termone$ is 
in the form $\appTRS(\termfour,\termfive)$; we have three cases:
\begin{varitemize}
\item
  $\termfour$ is semi-canonical. Then, we get what we want by induction hypothesis.
\item
  $\termfour$ is in $\TRSWconterms$ and has the form $\constr{\varone}{\lambdaone}(\termone_1,\ldots,\termone_m)$.
  Then, $n=0$ and $\termone$ is itself canonical.
\item
  $\termfour$ is in $\TRSWconterms$ and has the form $\cappTRSW(\termsix,\termseven)$. Then
  $$
  \termone=\appTRS(\cappTRSW(\termsix,\termseven),\termfive)\rewrTRSW\appTRS(\appTRS(\termsix,\termseven),\termfive).
  $$
  Apply now the induction hypothesis to $\appTRS(\termsix,\termseven)$ (since its length is
  strictly smaller than $\length{\termone}$).
\end{varitemize}
We can now proceed as in Lemma~\ref{lemma:TRStolam}, since whenever
$\termone$ rewrites to $\termtwo$ by one of the ordinary rules, $\termtwo$ is 
semi-canonical.
\qed
\begin{lem}\label{lemma:TRSWnormalform}
A canonical term $\termone\in\TRSWterms$ is in normal form iff $\TRSWtolambda{\termone}$ is in
normal form.
\end{lem}
\proof
We first prove that any canonical normal form $\termone$ can be written
as $\constr{\varone}{\lambdaone}(\termone_1,\ldots,\termone_n)$, where
$\termone_1,\ldots,\termone_n\in\TRSWconterms$. We proceed by induction
on $\termone$:
\begin{varitemize}
\item
  If $\termone=\constr{\varone}{\lambdaone}(\termone_1,\ldots,\termone_n)$,
  then the thesis holds.
\item
  If $\termone=\appTRS(\termtwo,\termthree)$, then $\termtwo$ is canonical
  and in normal form, hence in the form $\constr{\varone}{\lambdaone}(\termone_1,\ldots,\termone_n)$
  by induction hypothesis. As a consequence, $\termone$ is not a normal
  form, which is a contradiction.
\end{varitemize}
We can now prove the statement of the lemma, by distinguishing two cases:
\begin{varitemize}
\item
  If $\termone=\constr{\varone}{\lambdaone}(\termone_1,\ldots,\termone_n)$,
  where $\termone_1,\ldots,\termone_n\in\TRSWconterms$, then
  $\termone$ is in normal form and $\TRSWtolambda{\termone}$ is an
  abstraction, hence a normal form.
\item
  If $\termone=\appTRS(\termtwo,\termthree)$, then $\termone$ cannot be a normal form,
  since $\termtwo$ is canonical and in normal form and, as a consequence,
  it can be written as $\constr{\varone}{\lambdaone}(\termone_1,\ldots,\termone_n)$.
\end{varitemize}
This concludes the proof.
\qed
Observe that this property holds only if $\termone$ is canonical: a non-canonical term
can reduce to another one (canonical or not) even if the underlying $\lambda$-term
is a normal form.
\begin{lem}\label{lemma:lamtoTRSW}
If $\lambdaone\rewrlambdah\lambdatwo$, $\termone$ is canonical and $\TRSWtolambda{\termone}=\lambdaone$,
then $\termone\rewrTRSW\termtwo$, where $\TRSWtolambda{\termtwo}=\lambdatwo$ and 
$\plength{\termtwo}{\appTRS}+1\geq\plength{\termone}{\appTRS}$.
\end{lem}
\proof
Analogous to the one of Lemma~\ref{lemma:lamtoTRS} for the first part of the statement.
For the bound on the number of $\appTRS$, argument similarly to the proof
of Lemma~\ref{lemma:simul}.
\qed
The slight mismatch between call-by-name reduction in $\Lambdaterms$ and
reduction in $\TRSW$ is anyway harmless globally. As we now show, the total number of reduction
steps in $\TRSW$ is at most two times as large as the total number of call-by-name reduction
steps in $\Lambdaterms$.
\begin{thm}[Term Reducibility]
Let $\lambdaone\in\Lambdaterms$ be a closed term. The following
two conditions are equivalent:
\begin{enumerate}[\em 1.]
\item
  $\lambdaone\rewrlambdah^n\lambdatwo$ where $\lambdatwo$ is in normal form;
\item
  $\LambdatoTRSW{\lambdaone}\rewrTRSW^m\termone$ where
  $\TRSWtolambda{\termone}=\lambdatwo$ and $\termone$ is in normal form.
\end{enumerate}
Moreover $n\leq m\leq 2n$.
\end{thm}
\proof
Suppose $\lambdaone\rewrlambdah^n\lambdatwo$, where $\lambdatwo$ is in normal form.
$\lambdaone$ is closed and, by Lemma~\ref{lemma:closedcanonical}, $\LambdatoTRSW{\lambdaone}$
is canonical. By iterating over Lemma~\ref{lemma:simul} and Lemma~\ref{lemma:lamtoTRSW},
we obtain the existence of a term $\termone$ such that $\TRSWtolambda{\termone}=\termtwo$,
$\termone$ is in normal form and $\LambdatoTRSW{\lambdaone}\rewrTRSW^m\termone$, where $m\geq n$ and
$$
\plength{\termone}{\appTRS}-\plength{\LambdatoTRSW{\lambdaone}}{\appTRS}\geq (m-n)-n.
$$
Since $\plength{\termone}{\appTRS}=0$ ($\termone$ is in normal form), $m\leq 2n$.
If $\LambdatoTRSW{\lambdaone}\rewrTRSW^m\termone$ where
$\TRSWtolambda{\termone}=\lambdatwo$ and $\termone$ is in normal form, then
by iterating over Lemma~\ref{lemma:simul} we obtain that $\lambdaone\rewrlambdah^n\lambdatwo$
where $n\leq m\leq 2n$ and $\lambdatwo$ is in normal form.
\qed
$\GRSW$ is the graph rewrite system corresponding to $\TRSW$, in the sense of
Section~\ref{Sect:GraphRep}. Exactly as for the call-by-value case, computing the normal
form of (the graph representation of) any term takes time polynomial in the
number of reduction steps to normal form:
\begin{thm}\label{thm:mainw}
There is a polynomial $p:\N^2\rightarrow\N$ such that for every $\lambda$-term $\lambdaone$,
the normal form of $\TRStoGRSW{\LambdatoTRSW{\lambdaone}}$ can be computed in time at most 
$p(|\lambdaone|,\Timew{\lambdaone})$.
\end{thm}
On the other hand, we cannot hope to \emph{directly} reuse the results in Section~\ref{Sect:CTR2L}
when proving the existence of an embedding of OCRSs into weak call-by-name 
$\lambda$-calculus: the same $\lambda$-term can have distinct normal forms in the two cases. 
It is widely known, however, that a continuation-passing translation
can be used to simulate call-by-value reduction by call-by-name reduction~\cite{Plotkin75tcs}.
The only missing tale is about the relative performances: do terms obtained via the CPS 
translation reduce (in call-by-name) to their normal forms in a number of
steps which is \emph{comparable} to the number of (call-by-value) steps to normal form
for the original terms? We conjecture the answer is ``yes'', but we leave the task of proving that
to a future work.
\section{Conclusions}
We have shown that the most na\"\i{}ve cost models for weak call-by-value 
and call-by-name
$\lambda$-calculus
(each beta-reduction step has unitary cost)
and orthogonal constructor term rewriting (each rule application has unitary cost)
are linearly related. Since, in turn, this cost model for
$\lambda$-calculus is polynomially related to the actual cost of reducing
a $\lambda$-term on a Turing machine, 
the two machine models we considered are both \emph{reasonable} machines, when
endowed with their natural, intrinsic cost models (see also Gurevich's opus on Abstract State Machine
simulation ``at the same level of abstraction'', e.g.~\cite{Gurevich}).
This strong (the embeddings we consider are compositional), complexity-preserving
equivalence between a first-order and a higher-order model is the most important technical
result of the paper. 

Ongoing and future work includes the investigation of how much of this simulation could be
recovered either in a typed setting (see~\cite{SplawskiU99} for some of the difficulties),
or in the case of $\lambda$-calculus with strong reduction, where we reduce under an abstraction. 
Novel techniques have to be developed, since the analysis of the present 
paper cannot be easily extended to these cases.

\bibliographystyle{alpha}
\bibliography{wie}

\newcommand{\etalchar}[1]{$^{#1}$}
\begin{thebibliography}{BEG{\etalchar{+}}86}

\bibitem[BC92]{Bellantoni92CC}
Stephen Bellantoni and Stephen Cook.
\newblock A new recursion-theoretic characterization of the polytime functions.
\newblock {\em Computational Complexity}, 2:97--110, 1992.

\bibitem[BEG{\etalchar{+}}86]{TGRbarendregt}
H.~Barendregt, M.~Eekelen, J.~Glauert, J.~Kennaway, M.~Plasmeijer, and
  M.~Sleep.
\newblock Term graph rewriting.
\newblock In J.~de~Bakker, A.~Nijman, and P.~Treleaven, editors, {\em Volume
  II: Parallel Languages on PARLE: Parallel Architectures and Languages
  Europe}, pages 141--158. Springer-Verlag, 1986.

\bibitem[DLM08]{CIE2006}
Ugo Dal~Lago and Simone Martini.
\newblock The weak lambda-calculus as a reasonable machine.
\newblock {\em Theoretical Computer Science}, 398:32--50, 2008.

\bibitem[DLM09]{DLMicalp}
Ugo Dal~Lago and Simone Martini.
\newblock On constructor rewrite systems and the lambda-calculus.
\newblock In {\em Automata, Languages and Programming, 36th International
  Colloquium, Proceedings}, volume 5556 of {\em LNCS}, pages 163--174.
  Springer, 2009.

\bibitem[DLM10]{DLM09}
Ugo Dal~Lago and Simone Martini.
\newblock Derivational complexity is an invariant cost model.
\newblock In {\em Foundational and Practical Aspects of Resource Analysis,
  First International Workshop, Proceedings}, volume 6324 of {\em LNCS}, pages
  88--101. Springer, 2010.

\bibitem[Gir98]{Girard98ic}
Jean-Yves Girard.
\newblock Light linear logic.
\newblock {\em Information and Computation}, 143(2):175--204, 1998.

\bibitem[Gur01]{Gurevich}
Yuri Gurevich.
\newblock The sequential {ASM} thesis.
\newblock In {\em Current trends in theoretical computer science}, pages
  363--392. World Scientific, 2001.

\bibitem[Jon87]{PJ87}
Simon~Peyton Jones.
\newblock {\em The Implementation of Functional Programming Languages}.
\newblock Prentice Hall, 1987.

\bibitem[Lei95]{Leivant95RRI}
Daniel Leivant.
\newblock Ramified recurrence and computational complexity {I}: word recurrence
  and poly-time.
\newblock In {\em Feasible Mathematics II}, pages 320--343. Birkh\"auser, 1995.

\bibitem[MM00]{Marion00}
Jean-Yves Marion and Jean-Yves Moyen.
\newblock Efficient first order functional program interpreter with time bound
  certifications.
\newblock In {\em Logic for Programming and Automated Reasoning, 7th
  International Conference, Proceedings}, volume 1955 of {\em LNCS}, pages
  25--42. Springer, 2000.

\bibitem[Par90]{Parigot89CSL}
Michel Parigot.
\newblock On the representation of data in lambda-calculus.
\newblock In {\em Computer Science Logic, 3rd International Workshop,
  Proceedings}, volume 440 of {\em LNCS}, pages 309--321. Springer, 1990.

\bibitem[Plo75]{Plotkin75tcs}
Gordon~D. Plotkin.
\newblock Call-by-name, call-by-value and the lambda-calculus.
\newblock {\em Theoretical Computer Science}, 1(2):125--159, 1975.

\bibitem[Plu90]{Plump90ggacs}
Detlef Plump.
\newblock Graph-reducible term rewriting systems.
\newblock In {\em Graph-Grammars and Their Application to Computer Science},
  volume 532 of {\em LNCS}, pages 622--636. Springer, 1990.

\bibitem[PR93]{ParigotRoziere93}
Michel Parigot and Paul Rozi{\`e}re.
\newblock Constant time reductions in lambda-caculus.
\newblock In {\em Mathematical Foundations of Computer Science 1993, 18th
  International Symposium, Proceedings}, volume 711 of {\em LNCS}, pages
  608--617. Springer, 1993.

\bibitem[SGM02]{Sands:Lambda02}
D.~Sands, J.~Gustavsson, and A.~Moran.
\newblock Lambda calculi and linear speedups.
\newblock In {\em The Essence of Computation: Complexity, Analysis,
  Transformation. Essays Dedicated to Neil D. Jones}, number 2566 in LNCS,
  pages 60--82. Springer, 2002.

\bibitem[SU99]{SplawskiU99}
Zdzislaw Splawski and Pawel Urzyczyn.
\newblock Type fixpoints: Iteration vs. recursion.
\newblock In {\em Functional Programming, 4th International Conference,
  Proceedings}, pages 102--113. ACM, 1999.

\bibitem[vEB90]{vanEmdeBoas90}
Peter van Emde~Boas.
\newblock Machine models and simulation.
\newblock In {\em Handbook of Theoretical Computer Science, Volume A:
  Algorithms and Complexity (A)}, pages 1--66. MIT Press, 1990.

\bibitem[Wad80]{Wadsworth80}
Christopher Wadsworth.
\newblock Some unusual $\lambda$-calculus numeral systems.
\newblock In J.P. Seldin and J.R. Hindley, editors, {\em To H.B. Curry: Essays
  on Combinatory Logic, Lambda Calculus and Formalism}. Academic Press, 1980.

\end{thebibliography}
\end{document}